\documentclass[useAMS, usenatbib]{mn2e}
\usepackage{times}
\usepackage{aas_macros, amsmath, amssymb, rotating, hyperref}

\title[On the detection of TeV $\gamma$-rays from GRB]{On the detection of TeV $\gamma$-rays from GRB with km$^3$ neutrino telescopes --- I. Muon event rate from single GRBs}
\author[T.L. Astraatmadja]{Tri L. Astraatmadja\thanks{E-mail:T.Astraatmadja@nikhef.nl}\\
Nikhef -- National Institute for Subatomic Physics, Science Park 105 1098 XG Amsterdam, The Netherlands\\
LION -- Leiden Institute of Physics, Leiden University, PO Box 9504 2300 RA Leiden, The Netherlands
}

\begin{document}
\date{}
\pagerange{\pageref{firstpage}--\pageref{lastpage}} \pubyear{2002}
\maketitle
\label{firstpage}
\begin{abstract}
This is a preliminary study to examine the prospect of detecting TeV photons from $\gamma$-ray bursts (GRB) using km-size neutrino telescopes, specifically for the ANTARES neutrino telescope. Although optimized to detect upgoing neutrino-induced muons, neutrino telescopes nevertheless have a potential to detect high-energy photons by detecting downgoing muons from the electromagnetic cascade induced by the interaction of TeV photons with the Earth's atmosphere. The photon energy spectrum of a GRB is modeled by a simple power law and is normalized by simple energy considerations. Taking into account the absorption of TeV photons by cosmic infrared backgrounds, an optical depth table calculated from a model by \cite{fin10} is used and the arriving number of photons on top of the Earth atmosphere is determined. Muon production in the atmosphere is determined by considering two main channels of muon production: Pion photoproduction and direct muon pair production. The muon energy loss during their traverse from the surface to the bottom of the sea is determined using the standard muon energy loss formula. Assuming different detector sizes, the number of detectable muons from single GRB events located at different redshifts and zenith distances is determined. The background is calculated assuming it consists primarily of cosmic ray-induced downgoing muons. The detection significance is calculated and it can be concluded that to obtain at least $3\sigma$ detection significance, a typical GRB has to be located at redshift $z \lesssim 0.07$ if the detector's muon effective area is $A^{\mu}_{\rm eff} \sim 10^{-2}\;{\rm km}^{2}$, or redshift $z \lesssim 0.15$, if the muon effective area is $A^{\mu}_{\rm eff} \sim 1\;{\rm km}^{2}$.
\end{abstract}


\section{Introduction}
A downgoing TeV $\gamma$-ray passing through the Earth's atmosphere will initiate an electromagnetic shower that contains a small number of muons. A kilometer-size neutrino telescope can detect these muons by detecting the Cherenkov photons produced from the interaction of these downgoing muons with the medium surrounding the telescope. The signals can then be reconstructed in order to obtain the photon's energy and its direction of arrival. This method of detection can, in principle, be applied to various TeV $\gamma$-ray sources but for reasons that will be discussed in the following, this paper concentrates on detecting TeV photons from $\gamma$-ray bursts (GRB). This paper is part of the theme of GRB detection by neutrino telescopes. Neutrino-induced muon event rates has been calculated in detail (e.g. \citealt{hal99, alv00, gup02, der03}), and this paper is complementary to them by providing the necessary calculations of photon-induced muon event rates.

Since the first detection of $\gamma$-ray burst (GRB) of cosmic origin \citep{kle73} in 1969, many advances has been made contributing to the study of GRBs (see \citealt{fis95a,fis95b,par00} for excellent reviews of the early days of GRB astronomy). Many challenges have been answered and problems solved, but many questions still remain and one of the outstanding problems in TeV $\gamma$-ray astronomy is the detection of TeV photons emitted from GRBs. Many authors have constructed models that predict the existence of TeV emission (e.g.\ \citealt{mes94,der00,pee05,asa07}), and there are no indications of a cut-off energy ever observed in any GRB spectrum. It is therefore possible that TeV photons could be produced and if detected they can provide constraints and clues of the GRB production mechanism. Not only their detection can give clues on the intrinsic properties of GRBs, but they can also provide constraints on the extragalactic infrared background that attenuate TeV photons during their travel from the source to the observers.

Observational evidence for higher-energy photons has accumulated in the past years. BATSE\footnote{Burst And Transient Source Experiment, \url{http://www.batse.msfc.nasa.gov/batse/}{}} observed several thousands GRBs with photon energies up to 300 keV \citep{pac99}, followed later on by EGRET\footnote{Energetic Gamma-Ray Experiment Telescope, \url{http://heasarc.gsfc.nasa.gov/docs/cgro/egret/}{}} (an instrument on board the {\it Compton Gamma Ray Observatory}\footnote{\url{http://heasarc.gsfc.nasa.gov/docs/cgro/}{}}) observations of GRBs with energies up to 200 MeV \citep{din95, sch92} and even a peculiar GRB that lasts for 90 minutes and energies up to 18 GeV \citep{hur94}. Most recently, \emph{Fermi}\footnote{\url{http://fermi.gsfc.nasa.gov/}{}} observed GRBs with energies up to $\epsilon_{\gamma} \simeq 10$ GeV \citep{abd09b}. 

Moving higher up the energy scale, attempts have been made to detect TeV components of GRBs. Using coordinates distributed by the BATSE Coordinates Distribution Network (BACODINE) and later on by the GRB Coordinates Network (GCN), the \emph{Whipple}\footnote{\url{http://www.sao.arizona.edu/FLWO/whipple.html}{}} collaboration has observed 9 BATSE GRBs and 7 other GRBs announced by GCN within minutes to hours after the burst time given by the alert \citep{con97, hor07}. No evidence of TeV emission were found but upper limits were reported. The MAGIC\footnote{Major Atmospheric Gamma-ray Imaging Cherenkov, \url{http://wwwmagic.mppmu.mpg.de/}{}} Telescope, using the same observation principle as \emph{Whipple}, observed 9 GRBs announced by GCN and found no evidence of TeV emission as well \citep{alb07}. 

So far the only indication of TeV emission were detections by the HEGRA\footnote{High-Energy-Gamma-Ray Astronomy, \url{http://www.mpi-hd.mpg.de/hfm/HEGRA/HEGRA.html}{}} AIROBICC\footnote{AIRshower Observation By angle Integrating Cherenkov Counters} detector who claimed $\gtrsim 16\;{\rm TeV}$ emission from GRB 920925c \citep{pad98}, the Milagrito\footnote{A prototype of the Milagro Gamma-ray Observatory, \url{http://www.lanl.gov/milagro/index.shtml}{}} collaboration \citep{atk00, atk03, atk05} who report detection at 0.1 TeV, and the GRAND\footnote{Gamma Ray Astrophysics at Notre Dame, \url{http://www.nd.edu/~grand/}{}} array \citep{poi03} at 0.01 TeV. These reports, however, have marginal detection significance ($\simeq 3\sigma$) but nevertheless provide a tantalizing glimpse of the possible extension of the GRB energy spectrum. 

Undersea or under-ice neutrino telescopes are far less sensitive to TeV photons than Imaging Atmospheric Cherenkov Telescopes (IACTs). But why it is argued that km-size Cherenkov detectors such as ANTARES\footnote{Astronomy with a Neutrino Telescope and Abyss environmental RESearch project, \url{http://antares.in2p3.fr}{}}, IceCube\footnote{\url{http://icecube.wisc.edu/}{}}, or the future KM3NeT\footnote{km$^3$ NEutrino Telescope, \url{http://www.km3net.org/}} have the potential to observe TeV photons from GRBs is not only because of its large collecting area but also because of its wide field of view and high duty cycle. Its field of view can reach up to $\simeq 3\:{\rm sr}$ above the horizon and operate at $\simeq 95\%$ duty factor. This means that the detector is taking data most of the time and will almost always be able to record events from any nearby GRB above the horizon which emits photons numerous enough to induce muons in sufficient number to be detectable. On the other hand, IACTs can only operate on clear, moonless nights, and their slow slewing rate is inadequate to observe GRBs which are transient by nature. Moreover, if we know the time and direction of the burst, we then have the knowledge of when and where to observe and this can greatly reduce the amount of background of downgoing muons induced by cosmic rays. This two-piece of information can be provided by the GRB Coordinates Network (GCN)\footnote{\url{http://gcn.gsfc.nasa.gov/}{}}, a system that distribute alert notices to its subscribers whenever any spacecraft that is part of this network detect a potential GRB \citep{bar00}. 

At present five satellites are part of this network: HETE\footnote{High Energy Transient Explorer, \url{http://space.mit.edu/HETE/}{}} \citep{ric03}, INTEGRAL\footnote{INTErnational Gamma-Ray Astrophysics Laboratory, \url{http://www.esa.int/esaMI/Integral/}{}} \citep{win03}, \emph{Swift}\footnote{\url{http://heasarc.nasa.gov/docs/swift/swiftsc.html}{}} \citep{geh04}, \emph{Fermi} \citep{moi08}, and AGILE\footnote{Astro‐rivelatore Gamma a Immagini LEggero (Light-Imaging Gamma Astrophysical Detector), \url{http://agile.rm.iasf.cnr.it/}{}} \citep{coc02}. By performing time and position correlation of detected events with those provided by GCN we can significantly reduce the background and increase the possibility of detection at higher significance.

Despite these potentials, detecting the TeV component of $\gamma$-ray bursts is not without pitfalls. One of the main problems that comes to mind is the attenuation of TeV photons by ambient IR photons in the universe. Along their path from the source to the Earth, TeV photons collide with ambient IR photons and annihilating themselves, creating pairs of electron--positron. The cross section for such process is well-known but measuring the accurate spectral density of cosmic IR photons at all redshifts is still the main problem. I will discuss this problem in more details in Section \ref{subsec:attenuation} by confronting current attenuation models with observational data and choose the best model. This attenuation will consequently limit our observations only to the nearest GRBs.

Another crucial problem is to calculate the number of muons produced from a $\gamma$-shower with a certain spectral index. Two muon production mechanisms are identified: photoproduction and direct muon-pair production. Both mechanisms are low-cross-section process and are efficient at different energy regimes. Both mechanisms will be discussed in more detail in Section \ref{Sec:muonprod} and I will provide the necessary formula to determine the number of muon produced from $\gamma$-showers. In calculating the observed muon flux at detector level, I also take into account the muon energy loss caused by their passage through seawater (Section \ref{subsec:muonpassage}). Using all this, the number of detectable muons for single GRB events at different redshifts are calculated. This is outlined in Section \ref{sec:muflux1grb}.

Previous works had addressed these problems. In a similar vein with this paper, \citet{alv99} have calculated the muon flux from GRBs at various redshifts and physical properties, but for the muon production only the photopion channel---which falls quickly with increasing energies---is considered. Recently, \citet{hal09} rectified this problem and discussed the various channels from which muons can be produced, including muon pair production. However they only consider nearby, local, TeV sources and neglect altogether extragalactic sources. Furthermore their results indicate that local TeV sources are too faint, even an IceCube-size neutrino telescope require $\sim10$ years of integration time before a detection with $3\sigma$ significance is obtained.

This paper is the first part of a series of papers that try to address these problems and examine the possibilities of detecting TeV photons from GRBs. This paper (Paper I) will outline the processes that take place during the emission of TeV photons up to the detection of muons at detector level and will also formulate the basic working equations. Paper II will present a calculation of the muon event rate from stacked GRB sources, made using the latest distribution functions of GRB world model, and Paper III will present a data analysis from the ANTARES neutrino telescope.

\section{The photon spectrum of a GRB}
\label{grbphotspec}
\subsection{Normalization of the photon spectrum}
The photon spectrum of a GRB occuring at redshift $z$ is assumed to be constant during the whole duration of the burst. The burst duration in the observer's frame is $\Delta t = (1 + z)\Delta t_*$ (throughout this paper, asterisks will be used to indicate terms in the source's frame while terms without asterisk are terms in the observer's frame). The photon spectrum $N(\epsilon)$ of a GRB is approximated by a broken but smoothly connected power law, known as the Band spectrum, which is a model based on BATSE observations of 54 GRB \citep{ban93}: 
\begin{equation}
   \label{eq:band}
   \begin{split}
   N(\epsilon) = &\; f_{\gamma} \left[H(\epsilon_{\rm bk} - \epsilon)\exp\left(-(b-a)\tfrac{\epsilon}{\epsilon_{\rm bk}}\right)\left(\frac{\epsilon}{\epsilon_{\rm bk}}\right)^{-(a+1)}\right.\\
   &+ \left.H(\epsilon - \epsilon_{\rm bk})\exp(a-b)\left(\frac{\epsilon}{\epsilon_{\rm bk}}\right)^{-(b+1)}\right]
   \;\text{TeV$^{-1}$ cm$^{-2}$ s$^{-1}$}, 
   \end{split}
\end{equation}
where $(a,b)$ are respectively the spectral indices of the power law in the low- and high-energy regime demarcated by the break energy $\epsilon_{\rm bk}$, and $f_{\gamma}$ is the normalization constant in unit of photons TeV$^{-1}$ cm$^{-2}$ s$^{-1}$. The function $H(x)$ is the Heaviside step function defined as $H(x) = 1$ for $x\geq 0$ and $H(x) = 0$ otherwise.

The break energy is related to the directly measurable peak energy $\epsilon_{\rm pk*}$, which is the energy in which the $\nu f_{\nu} \equiv \epsilon^2 N(\epsilon)$ peaks, through
\begin{equation}
\label{eq:E_pk}
\epsilon_{\rm bk} = \frac{b-a}{1-a}\epsilon_{\rm pk}.
\end{equation}

BATSE observations extend only to several hundreds keV and in some cases to several MeV, but subsequent observations by later satellites confirmed that the power law also extends to at least several GeV (e.g. \citealt{hur94, gon03, abd09b}). Based on this observational evidence, in this work we assume that this power law function also extends to the TeV regime.

The normalization constant $f_{\gamma}$ is calculated by relating the energy spectrum in Equation \ref{eq:band} to its instrinsic isotropic-equivalent bolometric luminosity $L^{\rm iso}_{\rm bol*}$:
\begin{equation}
   \label{normalization}
   L^{\rm iso}_{\rm bol*} = 4\pi r^2_c(z)(1 + z)\int^{\Delta t}_{0}dt\int^{\infty}_{0}d\epsilon N(\epsilon) \epsilon,
\end{equation}
in which $r_c(z)$ is its comoving distance at redshift $z$:
\begin{equation}
r_c(z) = \int^z_0 dz'(1 + z')\frac{dl}{dz'},
\end{equation}
where $dl/dz$ is the cosmological line element defined as
\begin{equation}\label{line_element}
\frac{dl}{dz} = \frac{c}{H_0}\frac{1}{(1 + z)\sqrt{\Omega_\Lambda + \Omega_m(1 + z)^3}},
\end{equation}
in which $c$ is the speed of light, $H_0 = 72\text{ km s$^{-1}$ Mpc$^{-1}$}$ is the Hubble constant at the present epoch, and $(\Omega_\Lambda, \Omega_m) = (0.742, 0.258)$ is the present dark energy and matter density in the universe in units of the critical energy density, $3H_0^2/8\pi G$. It is assumed that the GRB emission spectrum is constant during the whole burst duration. It is also important to note that $L^{\rm iso}_{\rm bol*}$ is an isotropic-equivalent luminosity which assume that the $\gamma$-ray emission is isotropic and is not beamed. The true, beamed, bolometric luminosity $L^{\rm true}_{\rm bol*}$ is related to $L^{\rm iso}_{\rm bol*}$ by 
\begin{equation}
L^{\rm true}_{\rm bol*} = (1 - \cos\theta_j)L^{\rm true}_{\rm bol*},
\end{equation}
where $\theta_j$ is the opening angle of the jet. The average value of the opening angle is $\langle\theta_j\rangle \sim 6^\circ$ \citep{ghi07}, making $L^{\rm true}_{\rm bol*} \sim 0.0055 L^{\rm true}_{\rm bol*}$.

The integration in Equation \ref{normalization} is solved by locking the spectral index $a$ to the typical value of $a = 0$ (\citealt{pre00, nat05}) and letting the other values as free parameters. Solving the integration this way, we can obtain the photon flux $f_{\gamma}$:  
\begin{equation}
   f_{\gamma} = \frac{L^{\rm iso}_{\rm bol*}}{4\pi r^2_c(z)\Delta t_{*} \epsilon^2_{\rm bk*}\lambda_{\rm bol}}, 
\end{equation}
in which $\epsilon_{\rm bk*} = \epsilon_{\rm bk}(1 + z)$ is the break energy in the source's frame and $\lambda_{\rm bol}$ is a bolometric correction to the flux, which is the result of the integration in energy. To avoid a divergent flux in the integration, we do not integrate it to infinite energy but instead cut the spectrum off at maximum energy $\epsilon_{\rm max*} = 300\;{\rm TeV}$. At the moment the upper cutoff of the photon spectrum is not known, and in fact the taking of $300\;{\rm TeV}$ as the limit of the integration is quite arbitrary. Taking this in mind, the value of $\lambda_{\rm bol}$ is then

\begin{equation}
\label{eq:K}
\lambda_{\rm bol} = 
   \begin{cases} 
      -\frac{1}{b}\exp(-b) + \frac{1}{b} + \frac{\exp(-b)}{1 - b}\left[\left(\frac{\epsilon_{\rm max*}}{\epsilon_{\rm bk*}}\right)^{1 - b} - 1\right], & \text{for } b\neq 1\\
      -\frac{1}{b}\exp(-b) + \frac{1}{b} + \exp(-b)\ln\left(\frac{\epsilon_{\rm max*}}{\epsilon_{\rm bk*}}\right), & \text{for } b=1.
   \end{cases}
\end{equation}
Thus given $(L^{\rm iso}_{\rm bol*}, z, b, \Delta t_{*}, \epsilon_{\rm bk*})$ as parameters, we can construct the photon spectrum of any GRB.

\subsection{Photon absorption by ambient IR-photons}
\label{subsec:attenuation}
Along the path from the source to the Earth, $\gamma$-ray photons interact with extragalactic background light (EBL) through the $\gamma\gamma\rightarrow e^{+}e^{-}$ process, annihilating themselves and creating electron--positron pairs. A $\gamma$-ray photon with energy $\epsilon_\gamma$ can produce a pair of electron--positron if it impacts a background photon with threshold energy
\begin{equation}
\label{e_th}
\epsilon_{\rm th} = \frac{2\epsilon^2_{e}}{\epsilon_\gamma(1 - \mu_i)},
\end{equation}
where $\mu_i = \cos\theta_i$ is the angle of impact between the two photons. For head-on collisions, the wavelength of EBL photons which will interact with passing TeV photons is then
\begin{equation}
\lambda_{\rm EBL} \simeq \lambda_e\frac{\epsilon_\gamma}{2m_e c^2} = 2.4 \epsilon_\gamma\;\mu{\rm m},
\end{equation}
in which $\lambda_e = h/(m_ec)$ is the Compton wavelength for an electron and the input $\epsilon_\gamma$ is in TeV. We can see that TeV photons will interact strongly with IR photons in the EBL. 

The optical depth $\tau_{\gamma\gamma}(\epsilon_{\gamma},z)$ as a function of observed photon energy $\epsilon_{\gamma}$ and redshift $z$ can be calculated if we also know the differential number density of background photons $n(\epsilon_{\rm bg}, z)$ at energy $\epsilon_{\rm bg}$ and redshift $z$:
\begin{equation}
\label{tau}
\begin{split}
\tau_{\gamma\gamma}(\epsilon_{\gamma}, z) = &\; \tfrac{1}{2}\int^z_0 dz \frac{dl}{dz} \int^{1}_{-1} d\mu_i (1 - \mu_i) \\
   &\times \int^\infty_{\epsilon_{\rm min}} d\epsilon_{\rm bg} n(\epsilon_{\rm bg}, z)\sigma[\epsilon_{\gamma}(1 + z), \epsilon_{\rm bg}, \mu_i],
\end{split}
\end{equation}
in which $\epsilon_{\rm min} = \epsilon_{\rm th}(1 + z)^{-1}$, $\frac{dl}{dz}$ is the cosmological line element defined in Equation \ref{line_element}, and $\sigma(\epsilon_1, \epsilon_2, \mu_i)$ is the cross section of the $\gamma\gamma$ pair production \citep{bre34,gou67}:
\begin{equation}
\begin{split}
\sigma(\epsilon_1, \epsilon_2, \mu_i) = &\;\tfrac{3}{16}\sigma_T(1 - \beta^2)\\
   &\times\left[(3 - \beta^4)\ln\left(\frac{1 + \beta}{1 - \beta}\right) + 2\beta(\beta^2 - 2)\right],
\end{split}
\end{equation}
in which $\sigma_T$ is the Thomson cross-section and $\beta$ is the electron--positron velocity in the center-of-mass frame:
\begin{equation}
\beta = \sqrt{1 - \frac{2\epsilon^2_e}{\epsilon_1\epsilon_2(1 - \mu_i)}}.
\end{equation}

Directly observing EBL photons to obtain their photon distribution is difficult because of contamination issue from the instrument as well as from the zodiacal light. Source discrimination is also another issue: The Cosmic Infrared Background (CIB)---which is extragalactic in nature---must be discriminated from foreground objects such as discrete sources like stars and compact objects within the Galaxy, as well as diffuse sources such as light scattered and emitted by interplanetary dust and emission by interstellar dust (see \citet{hau01} for a review on this matter). 

There are many approaches in calculating the EBL photon density for all redshifts. One basic approach of doing it is by using ``backward models,'' in which we start from the existing galaxy count data and then model the luminosity evolution of these galaxies backward in time (e.g. \citealt{ste06}). Another approach is the ``forward evolution,'' performed by assuming a set of cosmological theory and semi-analytic merger-tree models of galaxy formation to determine the star formation history of the universe (e.g. \citealt{pri05, gil09}). Yet another approach is to focus on the properties and evolution of starlight, the primary source of CIB emission. This model integrates stellar formation rates and properties over time to obtain the amount of light emitted (e.g. \citealt{kne04, fin10}). 

\begin{figure}
\begin{center}
\includegraphics[width=84mm]{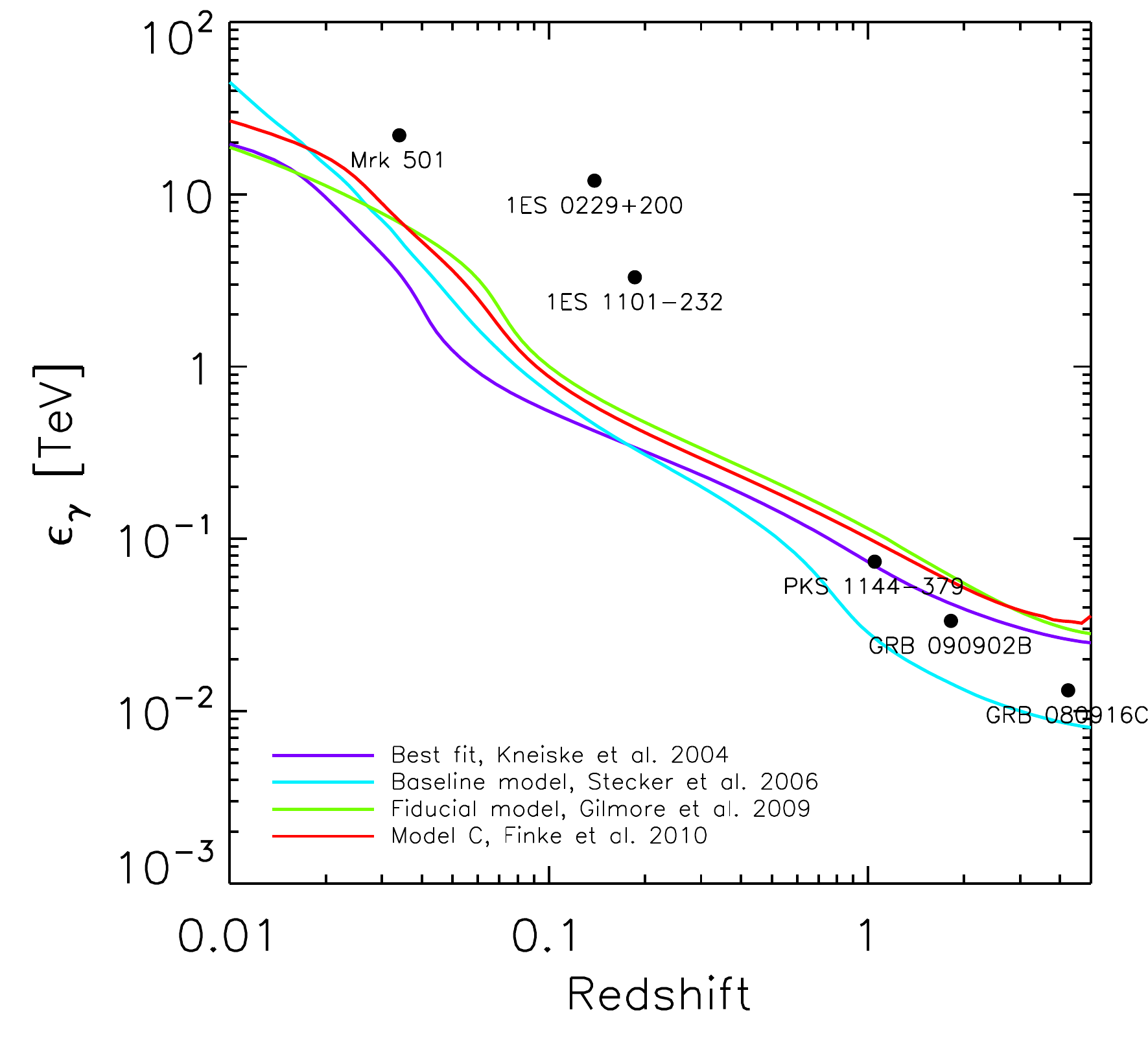}
\caption{A plot of the Fazio-Stecker Relationship \citep{faz70} for several attenuation models, as a function of redshift. Also shown are the redshifts and highest energy photons $\epsilon_{\rm max}$ of various objects observed by Atmospheric Cherenkov Telescopes and \emph{Fermi}-LAT \citep{fin09, abd10}.}
\label{fig:fsr}
\end{center}
\end{figure}

In this paper I consider three different attenuation models: The ``best-fit'' model of \citet{kne04}, the fiducial model of \citet{gil09}, and the recent ``Model C'' by \citet{fin10}. These models, along with the Baseline Model of \citet{ste06}, are compared in the plot of the Fazio-Stecker relation \citep{faz70} in Figure \ref{fig:fsr}. The Fazio-Stecker relation is the $(\epsilon_\gamma, z)$ value that gives $\tau_{\gamma\gamma} = 1$. This is interpreted to be the redshift at which the flux of photons of a given energy is attenuated by a factor $e$ and is called the $\gamma$-ray horizon. In this plot, for all models except those of \citet{ste06}, for redshift $\lesssim 5$ the universe is optically thin to photons with energy $\lesssim 20$ GeV. At very low redshifts however, the models are relatively consistent with each other, but the differences start to become apparent at $z\gtrsim 1$. The model calculated by \cite{ste06}, which predicts higher attenuation at higher redshifts, has in recent times contradicted MAGIC \citep{alb08} and \emph{Fermi} \citep{abd10} observations and thus can be ruled out with high confidence (furthermore, Figure $2$ in \citet{abd10} indicate that models by \citet{fin10, gil09, fra08} are the favorable ones) and will not be used in further calculations. 

\begin{figure}
\begin{center}
\includegraphics[width=84mm]{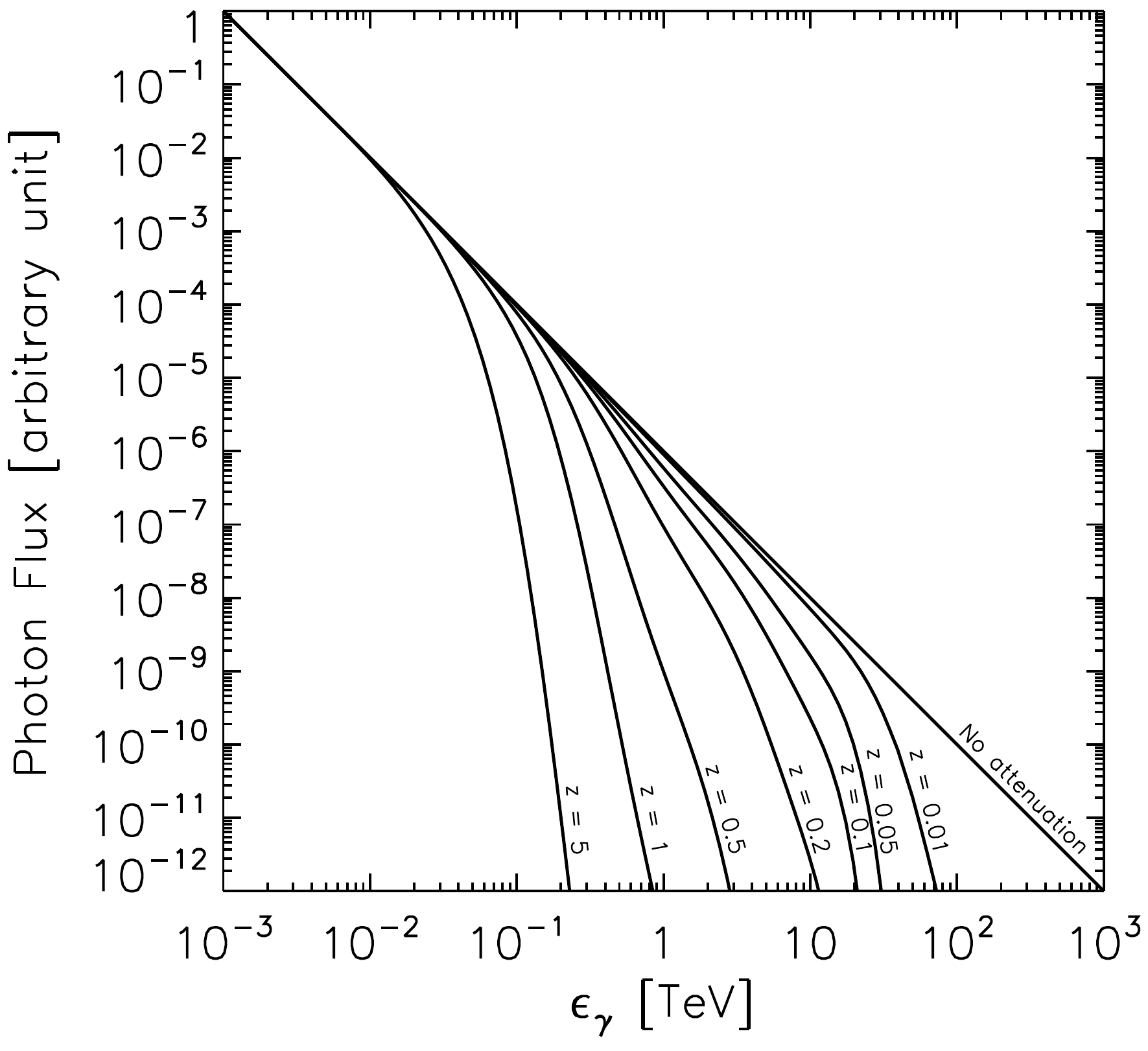}
\caption{An illustration of the effect of attenuation to a photon spectrum. Attenuation is calculated using the model by \citet{fin10}. The shape of the photon spectrum of a source located at redshifts indicated beside each curve is shown. Energies are in the observer frame of reference. The further a source is located, more attenuation is suffered by the highest energy photon. The curve is normalized to an arbitrary unit.}
\label{fig:photspec_attn}
\end{center}
\end{figure}

Thus, knowing the attenuation function, we can then estimate the total number of photons emitted from a GRB at redshift $z$ per unit energy arriving at the top of the Earth's atmosphere per unit area per unit time to be 
\begin{equation}
\label{eq:photspec_top}
\gamma_0(\epsilon_\gamma) = \gamma(\epsilon_\gamma,t=0) \equiv f_\gamma \left(\frac{\epsilon_{\gamma}}{\epsilon_{\rm bk}}\right)^{-(b+1)}e^{-\tau_{\gamma\gamma}(\epsilon_{\gamma}, z)}, 
\end{equation}
where $f_\gamma$ is as derived in Equation \ref{eq:K} and $\gamma(\epsilon,t)$ is the notation for the photon flux at slant depth $t$ in the atmosphere, as introduced in \citet{ros41}. Slant depth $t=0$ means the top of the atmosphere. In this equation only the high-energy part of Equation \ref{eq:band} is used, because this is precisely the concern of this study and henceforth this equation will be the working equation. 

To give an illustration of the effect of attenuation to a photon spectrum, the shape of the photon spectrum curve of several sources emitting at different redshifts is showed in Figure \ref{fig:photspec_attn}. As a comparison an unattenuated photon spectrum is also shown. The curves are normalized to an arbitrary unit. From the shape of the curves, the more distant the source is located, the more the photon spectrum curve is distorted due to attenuation effects. This imposes a limit on the number of TeV photons that we can observe.

\subsection{The cascade equation: Approximation A}
\label{subsec:cascade}
High energy photons interact with atoms in the atmosphere and initiate electromagnetic showers of particles that will cascade on their way through the atmosphere. Through materialization or Compton collision, pairs of electron-positron will be produced, which in turn emit additional photons by way of bremsstrahlung. At each step the number of particles increases but their average energy decreases \citep{ros41}. Nevertheless these secondary photons can also produce muons that can be detected by the detector array, and thus it is important to calculate the total number of photons produced in such a photon shower.

This problem of counting particles produced in electromagnetic showers can be solved if we consider only radiation phenomena and electron-pair production, which can be described by the asymptotic formula for complete screening. This solution is called Approximation A \citep{ros41} and allows us to calculate the photon flux at some depth $t$ in the atmosphere, given the initial photon energy spectrum. If the initial spectrum is in the form of a power law such as $\gamma(\epsilon) \propto \epsilon^{-(b+1)}$, then the resulting spectrum at depth $t$ is (\citealt{ros41,hal09})
\begin{equation}
   \label{eq:cascade}
   \begin{split}
   \gamma(\epsilon,t) =&\; \gamma(\epsilon,t=0)\frac{(\sigma_0 + \lambda_1)(\sigma_0 + \lambda_2)}{\lambda_2 - \lambda_1}\\
      &\times\left[\frac{\exp(\lambda_1 t)}{\sigma_0 + \lambda_1} - \frac{\exp(\lambda_2 t)}{\sigma_0 + \lambda_2}\right]
   \end{split}
\end{equation}

In this Equation as well as the in the following calculations, $t$ is the slant depth in units of radiation length (in the atmosphere, 1 radiation length equals $36.62\;{\rm g\;cm}^{-2}$), $\sigma_0 = 7/9$ is the probability per radiation length that an electron pair production will take place (in a case of complete screening), and $\lambda_{1,2}$ are the scale lengths factor of the shower growth and dissipation in the atmosphere. The formula to calculate $\lambda_{1,2}$ as a function of spectral index $b$, as well as its tabulation, is given in \citet{ros41}. For $b<1$, $\lambda_1$ is positive while for $b>1$, $\lambda_1$ is negative. This would mean that in the former case the shower would grow as it penetrates the atmosphere while in the latter it will dissipate. Thus for a general case of an arbitrary value of $b$, the photon flux can be decomposed into its spectrum at the top of the atmosphere and its scale factor at depth $t$, i.e 
\begin{equation}
   \gamma(\epsilon,t) = \gamma_0(\epsilon)\gamma_2(t).
\end{equation}

Particularly important is the case for $b=1$ since $\lambda_1 = 0$ and $\lambda_2 < 0$, and this would make the second exponential term in Equation \ref{eq:cascade} essentially zero after several radiation length, making the photon spectrum independent of depth:
\begin{equation}
   \gamma(\epsilon_\gamma, t) = 0.567\gamma(\epsilon_\gamma,t=0),
\end{equation}
where the photon spectrum at the top of the atmosphere $\gamma(\epsilon_\gamma, t=0)$ is as described in Equation \ref{eq:photspec_top}.

\section{Muon production in the atmosphere}
\label{Sec:muonprod}
High-energy $\gamma$-rays produce muons when they interact with the Earth's atmosphere. These muons will then traverse down to the bottom of the sea, producing Cherenkov light that can be detected by the detector array. This idea of detecting $\gamma$-induced showers by detecting the produced muons has been around for a long time. However, early calculations performed in the 1960s seem to indicate that $\gamma$-induced showers are muon-poor, having only less than 10\% the muon content of proton-induced showers \citep{sta85a}. These calculations are contradicted when muons were firmly detected at underground detectors, coming from the direction of Cygnus X-3 (e.g. \citealt{mar85}). Despite the low rates and weak signals, these detections raised the interest to build large-area detectors that can detect high-energy muons and thus operate as $\gamma$-ray observatory. \citet{sta85c} then identify two channels in which muons can be produced in $\gamma$ showers: photoproduction and direct muon-pair production. In photoproduction, muons are produced from the (semi)leptonic decay of pions or kaons produced by the interaction of high-energy photons with the atomic nucleus of the atmosphere. This is the most important channel to produce muons in the GeV regime. In direct muon-pair production, muons are created directly via the channel $\gamma + Z \rightarrow Z + \mu^{+} + \mu^{-}$, in which $Z$ is a nucleus of the atmosphere. Whereas muon production through photoproduction dies away with increasing energy, the cross section for muon-pair production increases with energy and thus muon-pair production is the dominant muon producing channel in the TeV regime.

In the following subsections we will describe the necessary formulation to calculate the muon flux generated in gamma-induced showers. For convenience, all units of length are converted into radiation lengths in the air $\lambda_{\rm rad}$, which is taken to be $37.1$ g cm$^{-2}$.

\subsection{Pion decay}
\label{subsec:pion_decay}
The interaction of high-energy photons with atomic nuclei in the atmosphere can produce pions through the reaction $\gamma + N \rightarrow \pi + X$ followed by leptonic decay of pions into a positive muon and a muon neutrino, or a negative muon and a muon antineutrino :
\begin{equation}
   \pi^\pm \rightarrow \mu^\pm + \nu_\mu(\overline{\nu}_\mu),
\end{equation}
with a probabilty of close to 100\% to occur. The formulation to calculate the muon spectrum from this channel has been calculated using the linear cascade equation and assuming a power-law photon spectrum with spectral index $b=1$ by \citet{dre89}, and its generalisation to an arbitrary spectral index by \citet{hal09}. 

For the case of $b \neq 1 $, this paper will closely follow that of \citet{hal09}, which begins by an ansatz that the differential pion spectrum in the atmosphere can be factorized as
\begin{equation}
   \pi(\epsilon,t) = \gamma(\epsilon,t=0)\pi_2(\epsilon,t),
\end{equation}
in which $\pi_2(\epsilon,t)$ can be split in two regimes: the high energy regime where pion interactions dominate over decay, and the low energy regime where pion interactions are neglected. The pion spectrum at high energy is
\begin{align}
   \label{eq:pispec_he}
   \pi^{\rm HE}_2(t) = & \left[\frac{\exp(\lambda_1t) - \exp(-t/\Lambda_{\pi})}{(\sigma_0 + \lambda_1)(\lambda_1 + \tfrac{1}{\Lambda_{\pi}})} - \frac{\exp(\lambda_2t) - \exp(-t/\Lambda_{\pi})}{(\sigma_0 + \lambda_2)(\lambda_2 + \tfrac{1}{\Lambda_{\pi}})}\right]\nonumber\\
      & \times\frac{z_{\gamma\pi}}{\lambda_{\gamma A}}\frac{(\sigma_0 + \lambda_1)(\sigma_0 + \lambda_2)}{\lambda_2 - \lambda_1},
\end{align} 
while the spectrum at low energy is
\begin{align}
   \label{eq:pispec_le}
   \pi^{\rm LE}_2(\epsilon,t) = & \frac{z_{\gamma\pi}}{\lambda_{\gamma A}}\frac{(\sigma_0 + \lambda_1)(\sigma_0 + \lambda_2)}{\lambda_2 - \lambda_1}\nonumber\\
      & \times\int^t_0 dt'\left(\frac{t'}{t}\right)^\delta\left[\frac{\exp(\lambda_1 t')}{\sigma_0 + \lambda_1} - \frac{\exp(\lambda_2 t')}{\sigma_0 + \lambda_2}\right],
\end{align}
in which $\delta = t/d_\pi$, where $d_\pi$ is the decay length
\begin{equation}
   d_\pi = \frac{\epsilon t\cos\theta}{\epsilon_\pi}, 
\end{equation}
here $\epsilon_\pi = 115\text{ GeV}$ is the pion decay energy constant.

The integral in Equation \ref{eq:pispec_le} can be expanded into series:
\begin{equation}
   \int^t_0 dt'\left(\frac{t'}{t}\right)^\delta\frac{\exp(\lambda_i t')}{\sigma_0 + \lambda_i} \approx \frac{1}{\sigma_0 + \lambda_i}\sum^{100}_{j=1}\frac{\lambda^{j-1}_i t^j}{(j-1)!(\delta + j)}.
\end{equation}
In Equation \ref{eq:pispec_he} and \ref{eq:pispec_le}, 
\begin{equation}
   \label{eq:Lambda_pi}
   \Lambda_\pi = 173 \text{ g cm}^{-2} = 4.66 \text{ radiation lengths}
\end{equation}
is the effective pion interaction length in the atmosphere, 
\begin{equation}
   z_{\gamma\pi} = \frac{\sigma_{\pi\pi}}{\sigma_{\gamma N}} = \tfrac{2}{3}
\end{equation}
is the ratio between cross sections $\sigma_{\gamma\rightarrow\pi}$ and $\sigma_{\gamma N}$, and
\begin{equation}
   \label{eq:Lambda_gamma_A}
   \lambda_{\gamma A} = 446.14 \text{ radiation lengths}
\end{equation}
is the interaction length of photons in atmospheric nuclei. These values are assumed to vary little for different spectral indices and energy.

Due to the unavailability of an analytical expression for both energy regime, taking a smooth transition from one regime to another is difficult. The pion spectrum at all energy regime is then
\begin{equation}
   \pi(\epsilon,t) = \gamma(\epsilon,t=0)\min\left[\pi^{\rm HE}_2(t),\pi^{\rm LE}_2(\epsilon,t)\right].
\end{equation}
The muon flux at the surface of the Earth can then be obtained by using standard 2-body decay kinematics, assuming no muon decay and energy loss in the atmosphere:
\begin{equation}
   \label{eq:muonflux_pion0}
   \frac{dN_\mu}{d\epsilon_\mu} = \int^{t_{\rm max}}_0 dt B_{\mu\pi}\int^{\epsilon/r}_\epsilon \frac{d\epsilon'}{(1 - r)\epsilon'}\frac{\pi(\epsilon',t)}{d_\pi(t)},
\end{equation}
in which $r = (m_\mu/m_\pi)^2$ and $B_{\mu\pi} = 1$ is the number of muons produced for each decaying pion. The maximum depth $t_{\rm max}$ is determined using 
\begin{equation}
   t_{\rm max} = \lambda_{e^+e^-}\ln\left[\frac{\epsilon_{\max}\langle x\rangle_{\gamma\rightarrow\mu}}{\epsilon}\right],
\end{equation}
where $\lambda_{e^+e^-} = 9/7$ is the electromagnetic cascade length and $\langle x\rangle_{\gamma\rightarrow\mu} = 0.25$ is the fraction of $\gamma$-ray energy that goes into the final muon for the case of pion decays.
   
For the special case of $b=1$, we calculate the muon spectrum using the formulation by \citet{dre89}: 
\begin{equation}
   \label{eq:muonflux_pion1}
   \frac{dN_\mu}{d\epsilon_\mu} = \gamma(\epsilon_\mu, t=0)\frac{\Lambda_\pi}{\lambda_{\gamma A}}z_{\gamma\pi}\frac{L_\gamma}{1 + (L_\gamma/H_\gamma){\epsilon_\mu}{\epsilon_\pi}\cos\theta},
\end{equation}
where
\begin{equation}
L_\gamma = \frac{1 - r^2}{2(1 - r)}\frac{t_{\rm max}}{\Lambda_\pi},\quad
H_\gamma = \frac{1 - r^3}{3(1 - r)}\left[1 + \ln\frac{t_{\rm max}}{\Lambda_\pi}\right].
\end{equation}
The constant terms $(\Lambda_{\pi}, z_{\gamma\pi}, \lambda_{\gamma A})$ in the Equations above are the same as in Equations \ref{eq:Lambda_pi}--\ref{eq:Lambda_gamma_A} 

\subsection{Direct muon-pair production}
\begin{figure}
\begin{center}
\includegraphics[width=5cm]{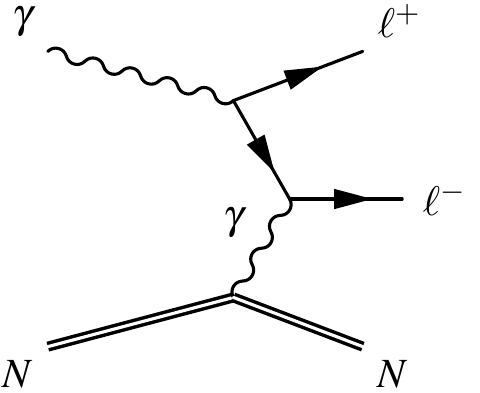}
\caption{Feynman diagram for lepton-pair production in the presence of a nucleus $N$}
\label{muonpairprod}
\end{center}
\end{figure}

The Feynman diagram for direct lepton-pair production $\gamma + N \rightarrow N + l^{+} + l^{-}$ is pictured in Figure \ref{muonpairprod}. This reaction occurs when an impacting photon interacts with a photon within the electric field of a nucleus, producing a pair of leptons. The second photon is necessary to maintain the conservation of 4-momentum, transferring the required momentum from the nucleus. Lepton-pair production is related to bremsstrahlung by a substitution rule and the calculation of the cross section can be done if we know how to calculate bremsstrahlung by electrons \citep{tsa74}. For the interaction of a photon with nuclear electrons to produce muon-pair, the photon energy threshold must then be
\begin{equation}
\epsilon_{\rm th} = \frac{2m_\mu}{m_e}\left(m_\mu + m_e\right) \simeq 43.9\:{\rm GeV},
\end{equation} 
where $m_e$ is the electron mass and $m_\mu$ is the muon mass.

To calculate an approximate formula of muon-pair production, what is usually done is taking the Bethe-Heitler result for electron-pair production \citep{bet34} and substitute the electron mass with that of muon. This generalization would not be correct, however, because the atomic form factor involved in the calculation must be integrated over the transferred momentum in which the upper limit is approximately the mass of the lepton involved \citep{hal09}. 

We will now discuss the necessary calculations to obtain the accurate formula for the cross section of muon-pair production.

The impacting photon energy will be fully shared by the resulting muon-pair according to
\begin{equation}
   \epsilon_{\gamma} = \epsilon^{+}_\mu + \epsilon^{-}_\mu,
\end{equation}
or in terms of fraction of photon energy:
\begin{equation}
   x_{+} = \frac{\epsilon^{+}_\mu}{\epsilon_\gamma}, \quad x_{-} = \frac{\epsilon^{-}_\mu}{\epsilon_\gamma}, \quad x_{+} + x_{-} = 1.
\end{equation}

To take into account the atomic and nuclear form factors, we need the differential cross section equation as a function of the momentum transfer. Since this work concerns very high-energy photons, we can use the ultrarelativistic approximation written as \citep{bet34}
\begin{equation}
   \label{eq:diff_cross_sec}
   \frac{d\sigma}{dx_{+}} = 4\alpha Z^2\left(r_0\frac{m_e}{m_\mu}\right)^2\left[\left(x^2_{+} + x^2_{-}\right)\Phi_1(\delta) + \frac{2}{3}x_{+}x_{-}\Phi_2(\delta)\right],
\end{equation}
where $alpha$ is the fine-structure constant, $Z$ is the charge of the nucleus---for the Earth's atmosphere $Z= 7.37$ \citep{ros52}, $r_0$ is the classical electron radius, and $\delta$ is the screening parameter equal to the necessary minimum momentum transfer from the nucleus:
\begin{equation}
\delta \simeq q_{\rm min} = \frac{m^2_\mu}{2\epsilon_\gamma x_{+}x_{-}}.
\end{equation}

The functions $\Phi_{1,2}$ are integrals of form factors over transferred momentum $q$. Whereas electron-pair production involves only the atomic form factors, in the case of muon-pair production it is also necessary to consider the nuclear form factors since the momentum involved is much larger than the inverse square of the atomic radius \citep{tsa74}. The functions $\Phi_{1,2}$ would then be
\begin{equation}
\label{eq:form_factor}
\Phi_{1,2}(\delta) = \int^{q_{\rm max}}_\delta \frac{dq}{q^3}\left[F_n(q) - F_a(q)\right]^2\psi_{1,2}(q,\delta),
\end{equation} 
where $F_n$ and $F_a$ are respectively the nuclear and atomic form factors and $\psi_{1,2}$ are the wave functions of the nucleus.

Equation \ref{eq:form_factor} has been solved with several assumptions. We take the solution of \citet{kel95} in which a single function $\Phi(\delta) = \Phi_1 = \Phi_2$ is used for the case of complete screening. By taking the effects of complete screening into account we consider the fact that atoms are essentially neutral at large distance. This is because the electric charge of the nucleus get ``screened'' by the atomic electrons, i.e. their field are canceled by opposite electric charge of the atomic electrons, reducing the effective charge according to distance and thus limiting the maximum distance at which photons can still interact. 

The contribution from inelastic form factors is also considered. This must also be taken into account since muon bremsstrahlung occurs on electrons bound in the atom and not on free electrons \citep{kel95}. 

Having considered both elastic and inelastic form factors, Equation \ref{eq:diff_cross_sec} then becomes
\begin{equation}
   \label{eq:diff_cross_sec2}
   \frac{d\sigma}{dx}\left(x,\epsilon_\gamma\right) = 4\alpha Z^2\left(r_0\frac{m_e}{m_\mu}\right)^2\left[1 - \frac{4}{3}x(1 - x)\right]\left[\Phi_{\rm el}(\delta) + \frac{1}{Z}\Phi_{\rm in}(\delta)\right].
\end{equation}
The elastic contribution $\Phi_{\rm el}(\delta)$ is in the form of
\begin{equation}
\Phi_{\rm el}(\delta) = \ln\left[\Phi_{\infty}\frac{1 + (D_ne^{1/2} - 2)\delta/m_\mu}{1 + BZ^{-1/3}e^{1/2}\delta/m_e}\right],
\end{equation}
where
\begin{equation*}
\Phi_{\infty} = \frac{BZ^{-1/3}}{D_n}\frac{m_\mu}{m_e}, \quad \delta = \frac{m^2_\mu}{2\epsilon_\gamma x (1 - x)}, \quad e^{1/2} = 1.6187\ldots
\end{equation*}
\begin{equation}
   \begin{array}{lll}
      B = 202.4 & \quad D_n = 1.49           & \quad \text{for Hydrogen, and}\\
      B = 183   & \quad D_n = 1.54A^{0.27}   & \quad \text{otherwise.}\\
   \end{array}
\end{equation}   
Here $A$ is the atomic number of the nuclei involved. For our case of the Earth's atmosphere, $A = 14.78$ \citep{ros52}.

The inelastic contribution $\Phi_{\rm in}(\delta)$ is
\begin{equation}
   \Phi_{\rm in} = \ln\left[\frac{m_\mu/\delta}{m_\mu\delta/m^2_e + e^{1/2}}\right] - \ln\left[1 + \frac{1}{B'Z^{-2/3}e^{1/2}\delta /m_e}\right],
\end{equation} where $B' = 1429$. We can see that the differential cross section is symmetric in $x_{+}$ and $x_{-}$, thus we can write \begin{equation*}
x_{+}x_{-} = x - x^2,
\end{equation*} where $x$ substitutes either $x_{+}$ or $x_{-}$ and the other becomes $(1 - x)$.

\begin{figure}
\begin{center}
\includegraphics[width=84mm]{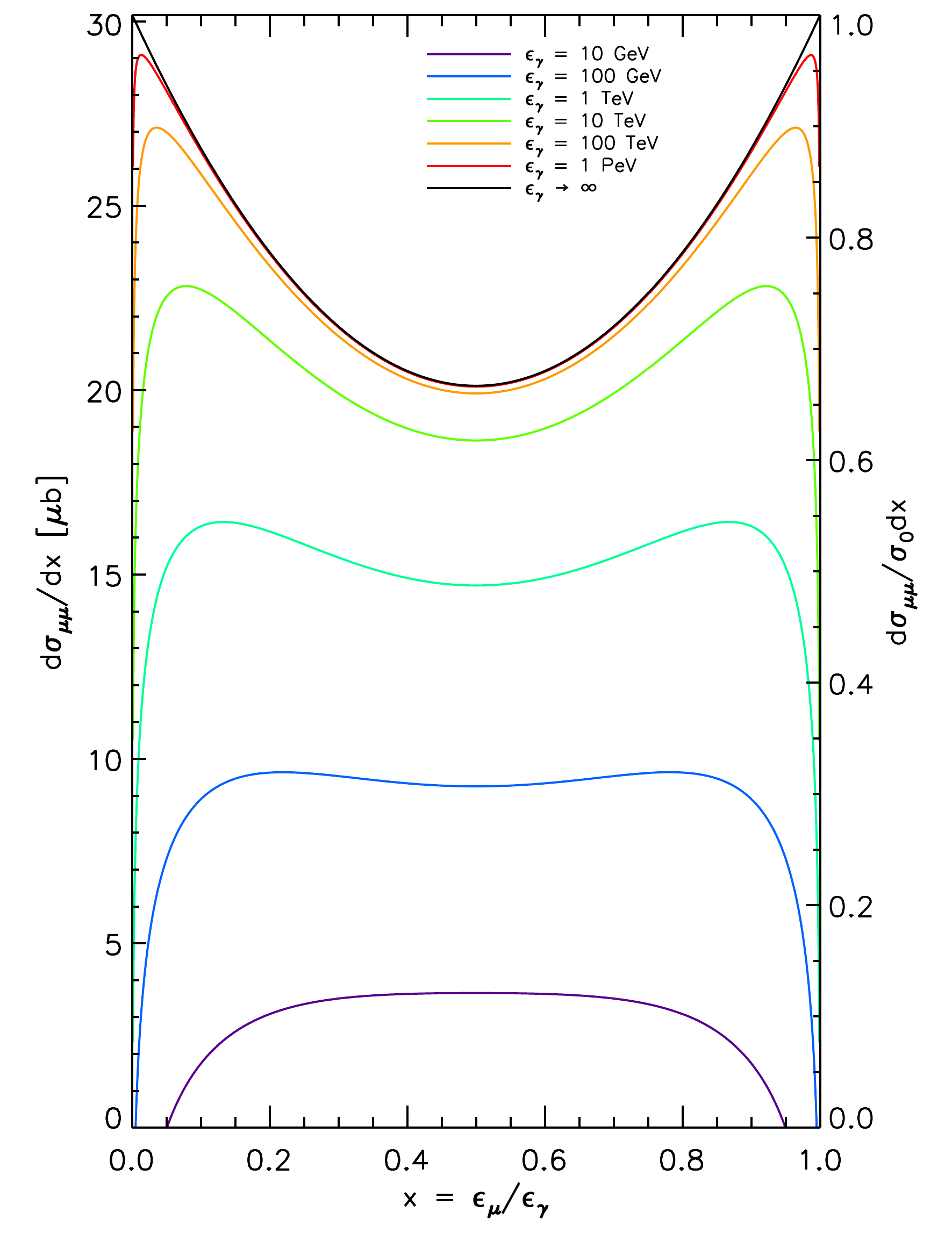}
\caption{Differential cross section of muon-pair production (Equation \ref{eq:diff_cross_sec2}) in the Earth's atmosphere for various values of impacting photon energy $\epsilon_\gamma$ as a function, as a function of $x = \epsilon_\mu/\epsilon_\gamma$ which is the ratio between the resulting muon energy and the photon energy. The atomic and mass number of the atmosphere is taken to be $(A, Z) = (14.78, 7.37)$.}
\label{fig:cross_sec_diff}
\end{center}
\end{figure}

\begin{figure}
\begin{center}
\includegraphics[width=84mm]{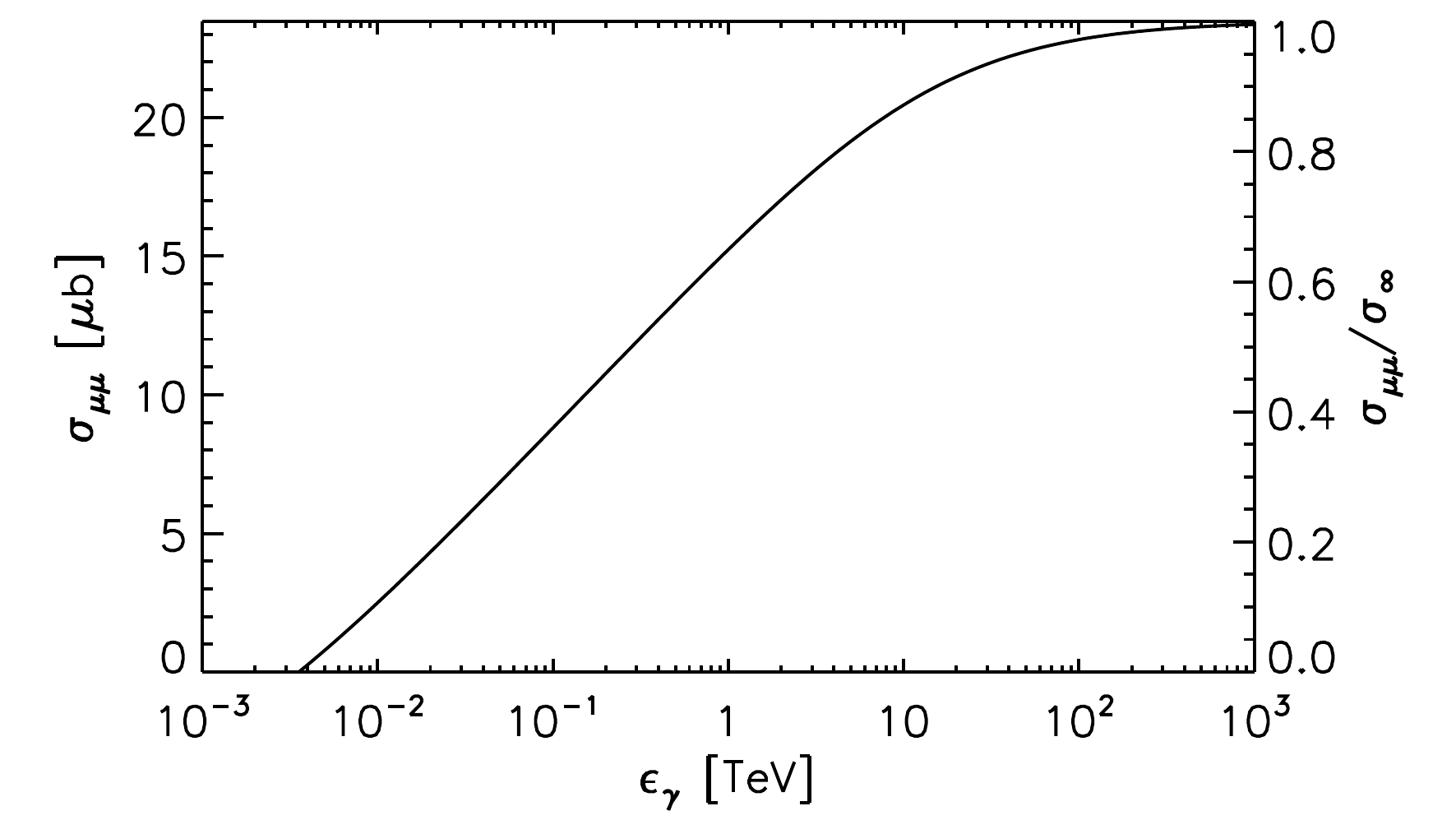}
\caption{Total cross section of the process $\gamma \rightarrow \mu^{+}\mu^{-}$ in the Earth's atmosphere as a function of impacting photon energy $\epsilon_\gamma$. Due to screening effect which limits the maximum distance in which high-energy photons can still interact with the nucleus, the cross section saturates for impacting photon energy $\epsilon_\gamma \gtrsim 10\:\text{TeV}$.} 
\label{cross_sec_total}
\end{center}
\end{figure}

In Figure \ref{fig:cross_sec_diff} Equation \ref{eq:diff_cross_sec2} for various values of photon energy $\epsilon_\gamma$ is shown. We can see that due to the ``screening'' effect the cross section does not increase indefinitely but saturates as $\epsilon_\gamma$ increases. I integrate the differential cross section over $x$ to obtain the total cross section as a function of photon energy and the result is shown in Figure \ref{cross_sec_total}. In the figure it is shown that saturation of the cross section occurs when the impacting photon energy $\epsilon_\gamma \approx 10$ TeV.

Using the cascade equation, we can calculate the muon-pair flux at sea level:
\begin{equation}
   \label{eq:muonflux_pair}
   \frac{dN_\mu}{d\epsilon_\mu} = 2\lambda_{\rm rad}\frac{N_{A}}{A}\gamma_0\left(\epsilon_\mu\right) \int^1_0 dx x^b\frac{d\sigma}{dx}\left(x,\frac{\epsilon_\mu}{x}\right)\int^{t_{\rm max}}_0 dt \gamma_2(t,b),
\end{equation}
where $N_A$ is the Avogadro number.

\subsection{Other channels of muon production}
A $\gamma$-shower can also produce kaons and the hadronic decay of kaons can produce a positive muon and a muon neutrino or a negative muon and a muon antineutrino:
\begin{equation}
   K^\pm \rightarrow \mu^\pm + \nu_\mu(\overline{\nu}_\mu).
\end{equation}
This reaction, however, has only $\sim 63.5\%$ chance of occuring \citep{gai90}. Furthermore, results from \citet{hal09} showed that the muon yield from kaon decays and other channels involving kaons can be neglected.

Positrons produced in $\gamma$-showers can also produce pairs of muon by interaction with an atomic electron through reaction $e^+ e^- \rightarrow \mu^+ + \mu^-$. However, cross section for this reaction is very small and peaked at $\sim 61$ GeV and falls rapidly with energy and is essentially zero for $\epsilon_\mu \gtrsim 700$ GeV \citep{hal09}. Thus this production channel can also be neglected altogether.

\subsection{Cosmic ray-induced muon background}
In order to calculate the detection significance of photon-induced muons, we need to know the amount of the background in our observation. In our case of photon-induced muons detection, the background consists of cosmic-ray induced muons. These muons are produced mainly through leptonic decay of pions, which is essentially the same channel discussed in Section \ref{subsec:pion_decay}. Leptonic decay of Kaons is also another channel of muon production albeit it is less important.

The energy spectrum of cosmic-ray induced muons, as a function energy and zenith distance, has already been parametrized by \citet{gai90} as
\begin{equation}
\label{muonbg}
\begin{split}
\frac{dN_\mu}{d\epsilon_\mu} \approx &\; 0.14E^{-2.7}\left[\frac{1}{1 + \tfrac{1.1\epsilon_\mu \cos\theta}{115 \text{GeV}}} \right.\\
   &\left.+ \frac{0.054}{1 + \tfrac{1.1\epsilon_\mu \cos\theta}{850 \text{GeV}}}\right]\text{ GeV$^{-1}$ cm$^{-2}$ s$^{-1}$ sr$^{-1}$}. 
\end{split}
\end{equation} 
This parametrization overestimates the actual measured muon flux for energies below 10 GeV because at that energy regime muon decay and muon energy loss become important factors (see Figure $6.1$ in \citealt{gai90}). However, this will not be our concern since this is far below the energy regime we are interested in, and Equation \ref{muonbg} fits perfectly well for high-energy regime. This equation estimates the muon flux at sea level, thus if we want to estimate the muon background at detector we have to apply the appropriate muon energy loss formula for seawater. We will discuss this later in Section \ref{subsec:muonpassage}.

\section{Passage of muons through seawater}
\label{subsec:muonpassage}
Upon traversing a medium, energetic muons lose their energy through ionization and radiative processes. This energy loss can be treated by taking the standard formula to calculate the average energy loss \citep{bar52}
\begin{equation}
   \label{eq:muon_energy_loss}
   -\frac{d\epsilon}{dx} = a(\epsilon) + b(\epsilon)\epsilon,
\end{equation}
in which $a(\epsilon)$ is the ionization contribution of the energy loss, while $b(\epsilon) = b_{p}(\epsilon) + b_{b}(\epsilon) + b_{n}(\epsilon)$ is the radiative contribution consisting of $e^{+}e^{-}$ pair production $(b_{p})$, bremsstrahlung $(b_{b})$, and photonuclear interaction $b_{n}$.

Here I take the approach of \citet{kli01} by splitting $a(\epsilon)$ into two separate processes, $a(\epsilon) = a_{c}(\epsilon) + a_{e}(\epsilon)$, where $a_{c}$ is the classical ionization process sufficiently described by the ``Bethe'' equation \citep{nak10} and $a_e$ is the $e$ diagrams for bremsstrahlung treated as part of an ionization process. $a_c$ can thus approximated by
\begin{equation}
   \label{eq:e_loss1}
   a_c(\epsilon) = a_{c_0} + a_{c_1}\ln\left(\frac{W_{\rm max}}{m_\mu}\right),\quad W_{\rm max} = \frac{\epsilon}{1 + \frac{m^2_\mu}{2m_e\epsilon}},
\end{equation}
in which $W_{\rm max}$ is the maximum transferable energy to the electron and $m_{\mu,e}$ are respectively the masses of muon and electron. The coefficients, in units of $(10^{-6}\;{\rm TeV}\;{\rm cm}^{2}\;{\rm g}^{-1})$, are $(a_{c_0}, a_{c_1}) = (2.106, 0.0950)$ for $\epsilon \leq 45\;{\rm GeV}$ and $(a_{c_0}, a_{c_1}) = (2.163, 0.0853)$ for $\epsilon > 45\;{\rm GeV}$. For $a_e$, a polynomial approximation is used:
\begin{equation}
   \label{eq:e_loss2}
   \begin{split}
   a_e(\epsilon) =&\; 3.54 + 3.785\ln \epsilon + 1.15\ln^2 \epsilon \\
      &+ 0.0615\ln^3 \epsilon\quad 10^{-9}\;{\rm TeV}\;{\rm cm}^2\;{\rm g}^{-1},
   \end{split}
\end{equation} 
where $\epsilon$ is in units of GeV.

The terms of $b$ are parametrized in a polynomial function in the form
\begin{equation}
   \label{eq:e_loss3}
b_i(\epsilon) = \sum^4_{j=0} b_{ij}\ln^j\epsilon,\quad\text{where }i=p,b,n.
\end{equation}
Here the energy input $\epsilon$ is also in units of GeV. The values of coefficients for $b_{ij}$ is already calculated by \citet{kli01} and is tabulated in their Table II. These formulations of energy loss are expected to still valid for $\epsilon_{\rm detector} = 30\;{\rm GeV}-5\;{\rm TeV}$ and slant depth $(3-12)$ km with errors up to $\pm(6-8)\%$ \citep{kli01}.

\begin{figure}
\begin{center}
\includegraphics[width=84mm]{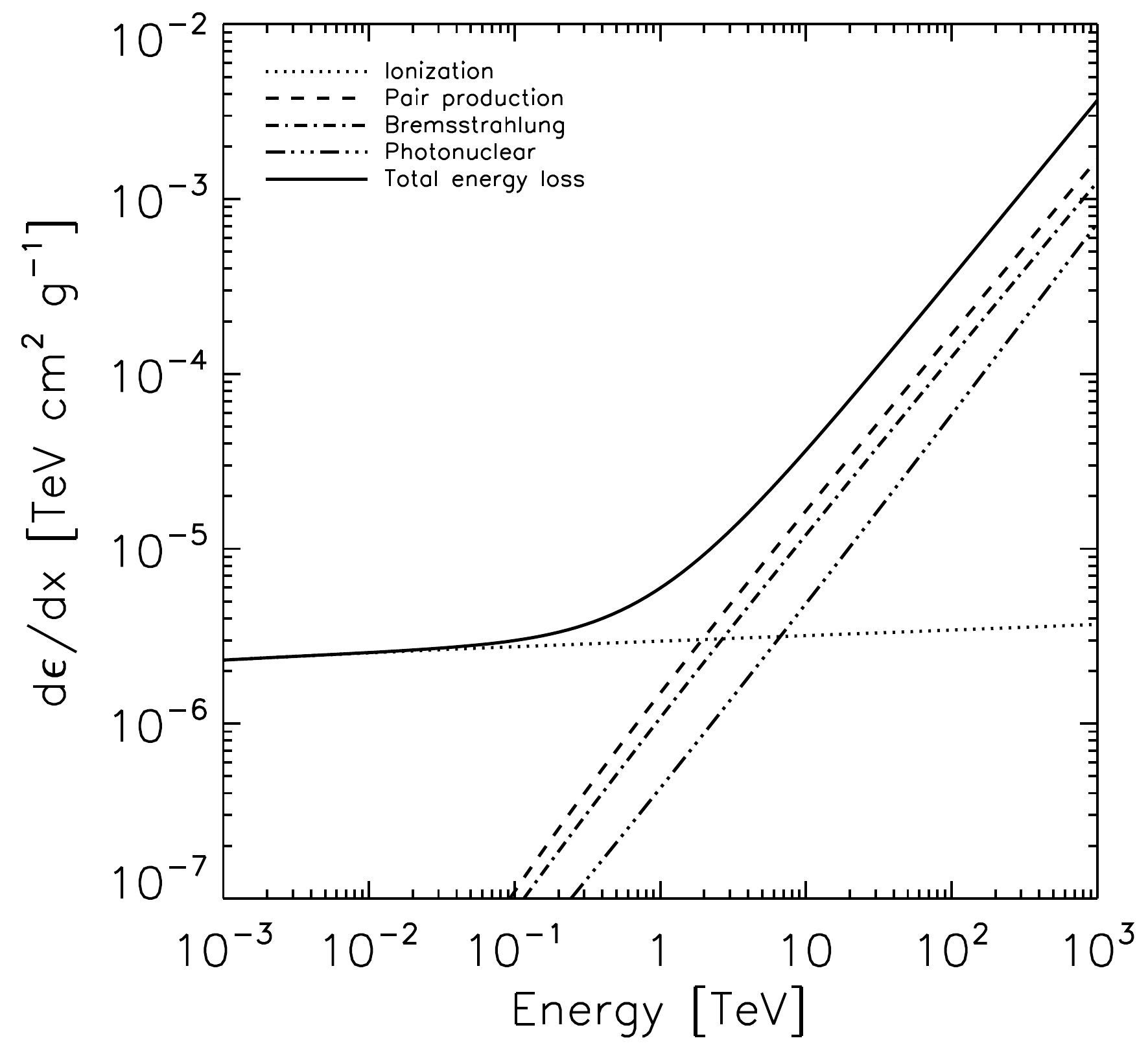}
\caption{\label{fig:muon_energy_loss}The muon energy loss in seawater as a function of energy, calculated from Equations \ref{eq:e_loss1} to \ref{eq:e_loss3}. The total energy loss (solid line) is decomposed into contributions from different processes, indicated in the legend. This Figure is made using the values of \citet{kli01}.}
\end{center}
\end{figure}
\begin{figure}
\begin{center}
\includegraphics[width=84mm]{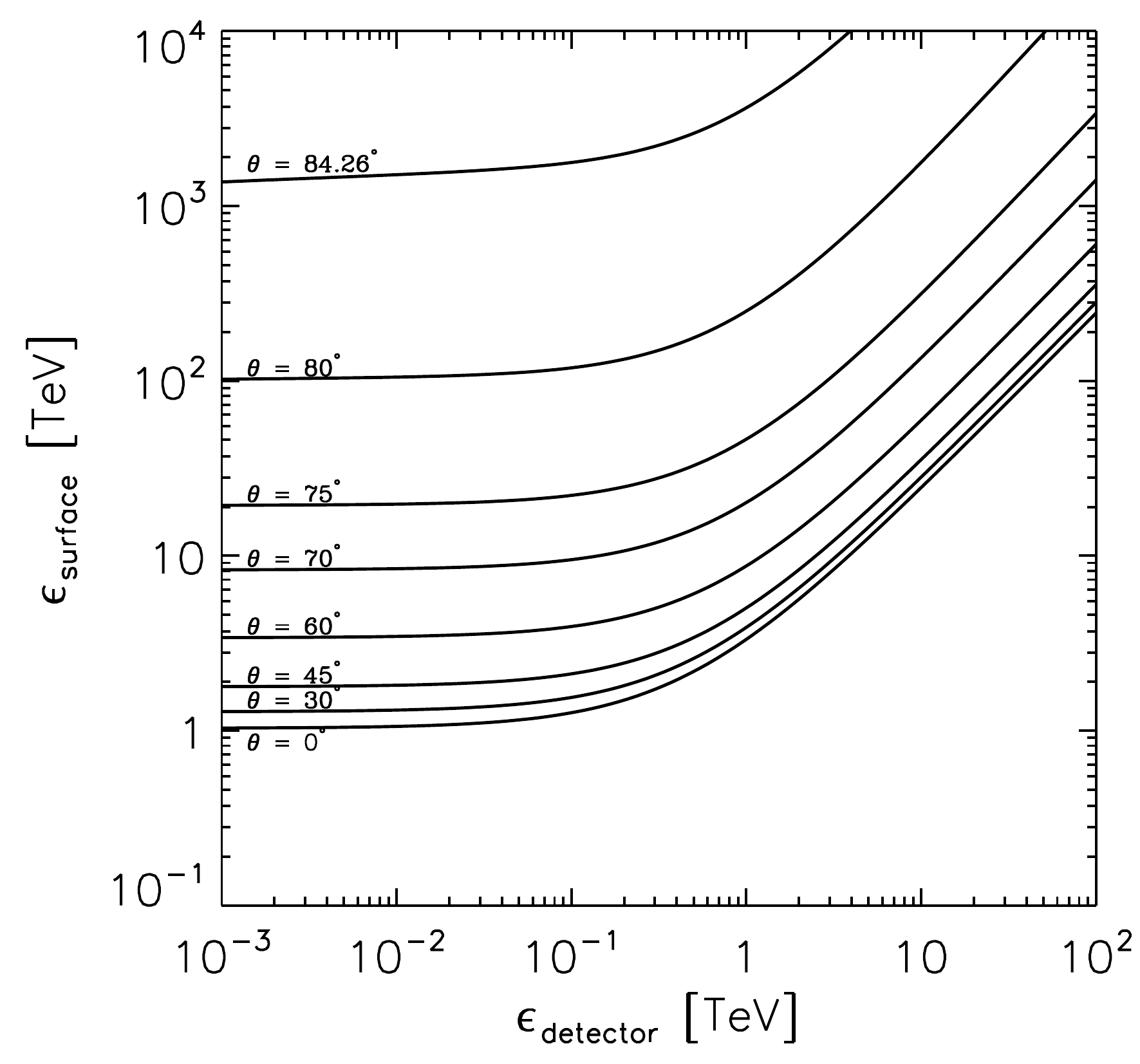}
\caption{\label{fig:muon_passage}The muon energy loss by passing a layer of sea water with vertical depth $d = 2475\text{ m}$ is pictured here in the form of muon energy at the surface of the sea $\epsilon_{\rm surface}$ as a function of muon energy at the detector level $\epsilon_{\rm detector}$. We plot the energy loss for different zenith distance $\theta$, thus the path length is $R=d/\cos\theta$.}
\end{center}
\end{figure}
Taking into account these contributions, the total muon energy loss in seawater as a function of energy is shown in Figure \ref{fig:muon_energy_loss}. In this figure we can see that at high energies radiative processes are more important than ionization. The critical energy at which the energy loss from ionization and radiative processes are equal can be calculated by solving $\epsilon_{\mu c} = a(\epsilon_{\mu c})/b(\epsilon_{\mu c})$. In the case of seawater this is $\epsilon_{\mu c} \sim 590\;{\rm GeV}$. Below this critical energy the dominant process is ionization while above this limit the radiative processes starts to dominate.

If we integrate Equation \ref{eq:muon_energy_loss} we can obtain the integral equation
\begin{equation}
\label{eq:integral_eq_energy_loss}
\int^{\epsilon_{\rm detector}}_{\epsilon_{\rm surface}}\frac{d\epsilon}{a(\epsilon) + b(\epsilon)\epsilon} + R = 0,
\end{equation}
in which $\epsilon_{\rm surface}$ is the energy at the surface of the sea and $\epsilon_{\rm detector}$ is the energy at detector level, located at slant depth $R = d/\cos\theta$ where $d$ is the vertical distance of the detector and $\theta$ is the zenith distance from which the source came. Solving this equation, we can obtain $\epsilon_{\rm surface}$ if $\epsilon_{\rm detector}$ is the input and vice versa. I solve Equation \ref{eq:integral_eq_energy_loss} to obtain $\epsilon_{\rm surface}$ as a function of $\epsilon_{\rm detector}$. The result for ANTARES depth of $d=2475\;{\rm m}$ below sea level is shown in Figure \ref{fig:muon_passage} for several slant depths.

The relation between $\epsilon_{\rm surface}$ as a function of $\epsilon_{\rm detector}$ is particularly useful to obtain the muon flux at detector level:
\begin{equation}
\label{eq:transport_muon}
\frac{dN}{d\epsilon_{\rm det}}(\epsilon_{\rm det}, R) = \frac{dN}{d\epsilon_{\rm sur}}(\epsilon_{\rm sur})\left.\frac{d\epsilon_{\rm sur}}{d\epsilon_{\rm det}}\right|_{\epsilon_{\rm det}, R}
\end{equation}

\section{Muon flux from single GRB}
\label{sec:muflux1grb}
\begin{figure*}
\begin{center}
\includegraphics[width=\textwidth]{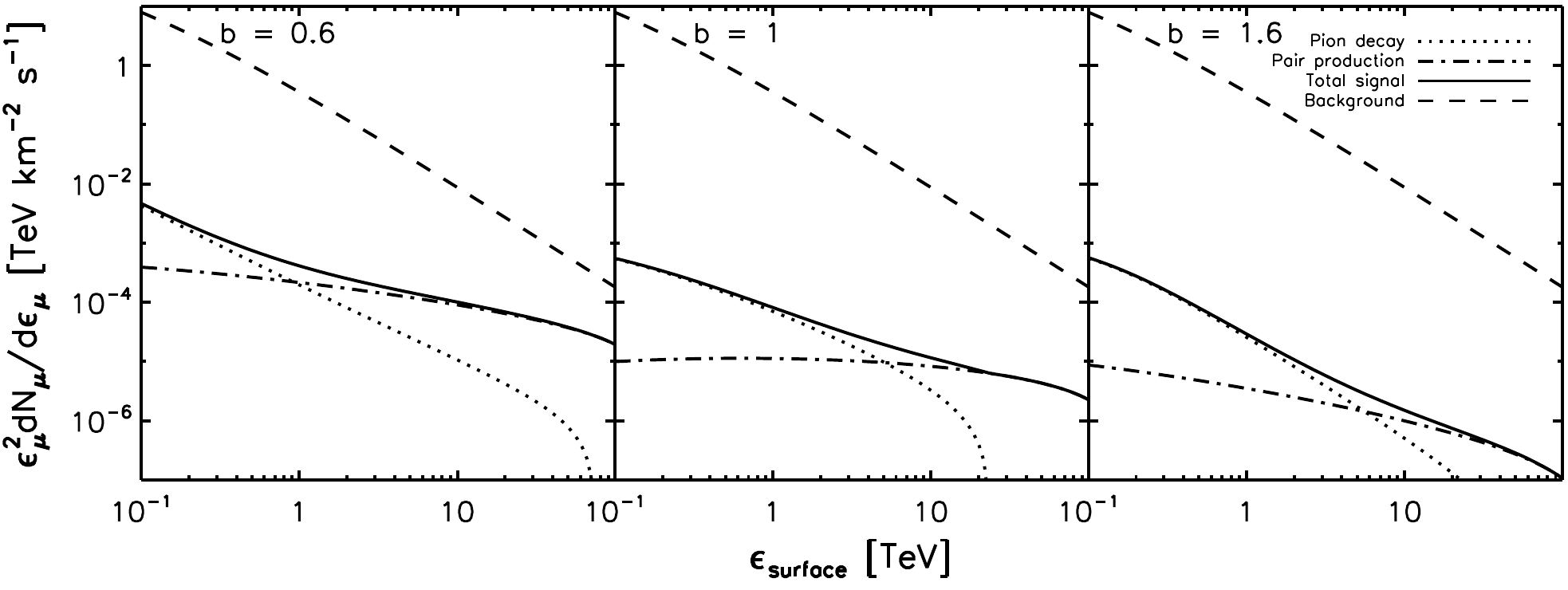}
\caption{The $\nu f_{\nu}$ spectrum of a fictive, unattenuated test source with fluence $f_\gamma = 10^{-1}$ TeV$^{-1}$ km$^{-2}$ s$^{-1}$ at 1 TeV. The spectrum is decomposed into its major contributing components: Pion decay and direct pair production. I calculate the spectrum for three different photon spectral index $s=(0.6,1,1.6)$ and compare it with a background spectrum of cosmic ray-induced muons calculated using Equation \ref{muonbg}, assuming that this fictive source is located at zenith distance $\theta = 30^{\circ}$. The search cone has an opening angle of $1^{\circ}$. The result is largely consistent with \citet{hal09}.} 
\label{fig:muonspec_test_source}
\end{center}
\end{figure*}

Once we know how to produce gamma ray-induced muons in the atmosphere and how they lose their energy in seawater, we are now in the position to calculate the muon yield both on the surface of the sea and at detector level. I first calculate muons produced from a fictive, unattenuated test source with fluence $f_\gamma = 10^{-1}$ TeV$^{-1}$ km$^{-2}$ s$^{-1}$ at 1 TeV. The source is a point source with negligible diameter, assumed to be located at zenith distance $\theta = 30^\circ$. The muon flux is calculated for three alternatives of spectral indices $b = (0.6,1,1.6)$ and cutoff energy at $\epsilon_{\rm max} = 300\;{\rm TeV}$. For the background estimation, the opening angle of the search cone is taken to be $\theta_{\rm cone} = 1^\circ$. The results are shown in Figure \ref{fig:muonspec_test_source} and compared to a background of cosmic ray-induced muon flux calculated for the same zenith distance.

These results are largely consistent with the results of \citet{hal09}. We can see that the dominant channel of muon-production at low energies is by pion decay. However the number of muons that can be created from this channel goes down with photon energy. At high energies, because the cross-section of the muon-pair production goes up with photon energy before reaching saturation point at $\epsilon_{\gamma}\gtrsim 10\;{\rm TeV}$, the dominant muon production mechanism is direct-pair production. 

Confident with consistency of the calculation, I proceed by calculating the muon flux for single GRB events located at different redshifts. Using Equation \ref{eq:photspec_top}, I calculate the photon flux arriving at the top of the atmosphere from GRBs with spectral indices $b=(0.5, 1, 1.25, 1.5)$, redshifts $z = (0.05, 0.1, 0.2, 0.5)$, and zenith distances $\cos\theta = (0.5, 1)$. A typical GRB power spectrum measured by BATSE is $b\simeq 1.25$ \citep{pre00}, however measurement inconsistencies has been reported and thus the shape of the spectral index at high energy is still debatable and might not be in the form of a simple power law (see e.g. \citealt{kan08} and \citealt{gon03}). Until this debate is clarified by \emph{Fermi}, it is reasonable also to asume a milder spectrum with index $b\simeq 1$. The other spectral indices, the shallower $b=0.5$ and the steeper $b=1.5$, while not entirely impossible nevertheless have a small possibility of occuring and is thus also considered to study their possibility of observing the muon signal.

Throughout the calculation, the values $\Delta t_* = 10\;{\rm s}$, $\epsilon_{\rm bk *} = (b - a)\epsilon_{\rm pk*}/(1 - a) = (b - 1)400\;{\rm keV}$, and $L^{\rm iso}_{\rm bol*} = 8.9\times 10^{52}\;{\rm erg}$ is used. The values taken for $\Delta t_*$, $\epsilon_{\rm pk*}$, and $L^{\rm iso}_{\rm bol*}$ are all the mean values determined from {\it Swift} results \citep{but07, but10}. After calculating the number of photons at the top of the atmosphere, the muon flux at the surface of the sea is then determined by means of Equation \ref{eq:muonflux_pion0} or \ref{eq:muonflux_pion1}---depending on the spectral index considered---and Equation \ref{eq:muonflux_pair}. The muon flux at the surface is then transformed to the muon flux at detector level by way of Equation \ref{eq:transport_muon}, and the corresponding energy at detector level is calculated by solving Equation \ref{eq:integral_eq_energy_loss}. 

\begin{figure}
\begin{center}
\includegraphics[width=84mm]{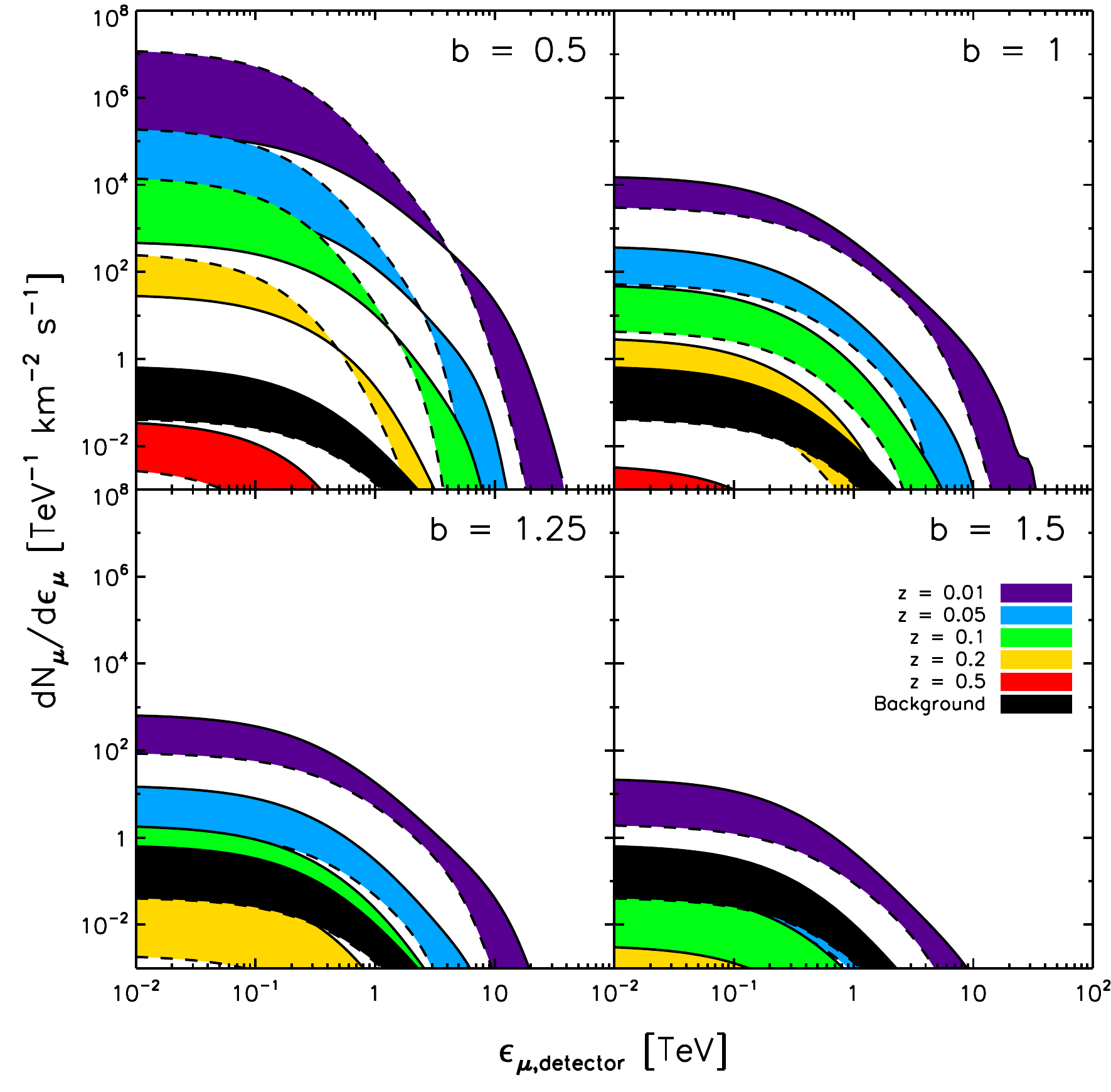}
\caption{The muon flux at detector depth (2475 m) for single GRBs emitted from different redshifts as indicated by the color coding on the legend. The color black is the background flux from cosmic ray-induced muons, calculated using Equation \ref{muonbg} assuming a search cone with an opening angle of $1^\circ$. For each color, the muon flux drawn by the dashed-line is the flux from zenith distance $\theta = 60^\circ$ while those drawn by the solid line is the flux straight from the zenith (i.e.\ $\theta = 0$). The filled-area then defines all the possible flux from all zenith distance between $\theta = 0$ and $\theta = 60^\circ$. Attenuation is determined by using a model by \citet{fin10}.}
\label{fig:muonspec_detlevel_finke10}
\end{center}
\end{figure}

The results of this series of calculations are shown in Figure \ref{fig:muonspec_detlevel_finke10} using the attenuation model by \citet{fin10}. One panel in each of these Figures plot the muon flux of GRBs for one spectral index. For each spectral index, the muon flux from GRBs at different redshifts is also shown and indicated with the color scheme shown in the legend. For each redshift, an area is drawn to show their dependence on zenith distance. The the borders of the area drawn for each redshifts are the the muon flux at zenith distance $\theta = 0$ (solid lines) and at $\theta = 60^{\circ}$ (dashed lines). Anything in between those two lines are then the amount of signals from any zenith distance between the borders. A background flux consisting of cosmic ray induced-muons calculated from Equation \ref{muonbg} is also shown for the same limit of zenith distances. In the Figure it is indicated by the black area. The search cone or the opening angle is taken to be $1^\circ$. I performed the same calculations for other attenuation models, but upon inspection of the numbers, results indicate that the magnitude of attenuation does not differ much for nearby universe, i.e. $z \lesssim 0.2$. Hence here only results calculated using the calculation by \citet{fin10} is shown.

For $b < 1$, we have a situation in which at low energy the muon yield at slanted angle is higher than the yield at vertical angle for the source located at the same distance. This is because according to the Cascade Equation discussed in Section \ref{subsec:cascade}, a photon shower with a shallow spectrum---i.e. having spectral index $b<1$---will instead grows instead of dissipating in the atmosphere. Because at slanted depth the photon passed through thicker layers of atmosphere, more photons are created in the cascade and thus have more chance to produce muons. At higher energies that situation is however no longer the case because the effectiveness of the pion decay channel in creating muons goes down with energy and the cross section of the muon-pair production channel reach saturation point. At slanted angle this limited number of high-energy muons then still have to pass a thicker layer of seawater than those that has to be traversed by muons coming from vertical direction.

\begin{figure}
\begin{center}
   \includegraphics[width=84mm]{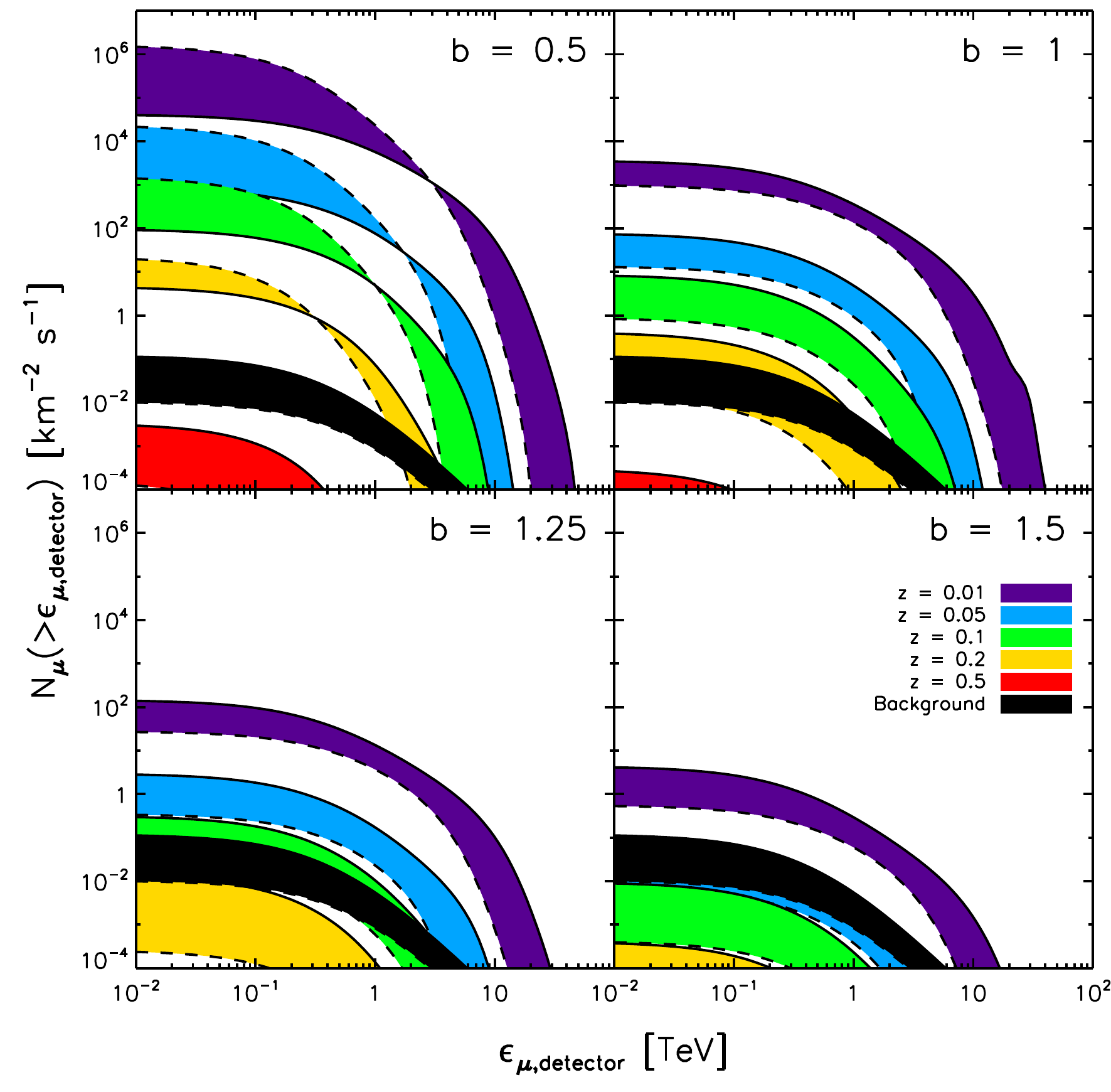}
   \caption{The expected number of muons with energies higher than the given muon energy $\epsilon_{\mu, {\rm detector}}$ at detector depth (2475 m) for single GRBs emitted from different redshift. The color black is the background flux. The same as in Figure \ref{fig:muonspec_detlevel_finke10}, dashed-lines indicate muon count from GRBs occuring at zenith distance $\theta = 60^{\circ}$, while solid lines are muon count from GRBs occuring at the zenith (i.e.\ $\theta = 0$). In this Figure, the EBL attenuation is calculated using the model by \citet{fin10}.}
   \label{fig:muoncount_detlevel_finke10}
\end{center}
\end{figure} 
\begin{figure}
\begin{center}
   \includegraphics[width=84mm]{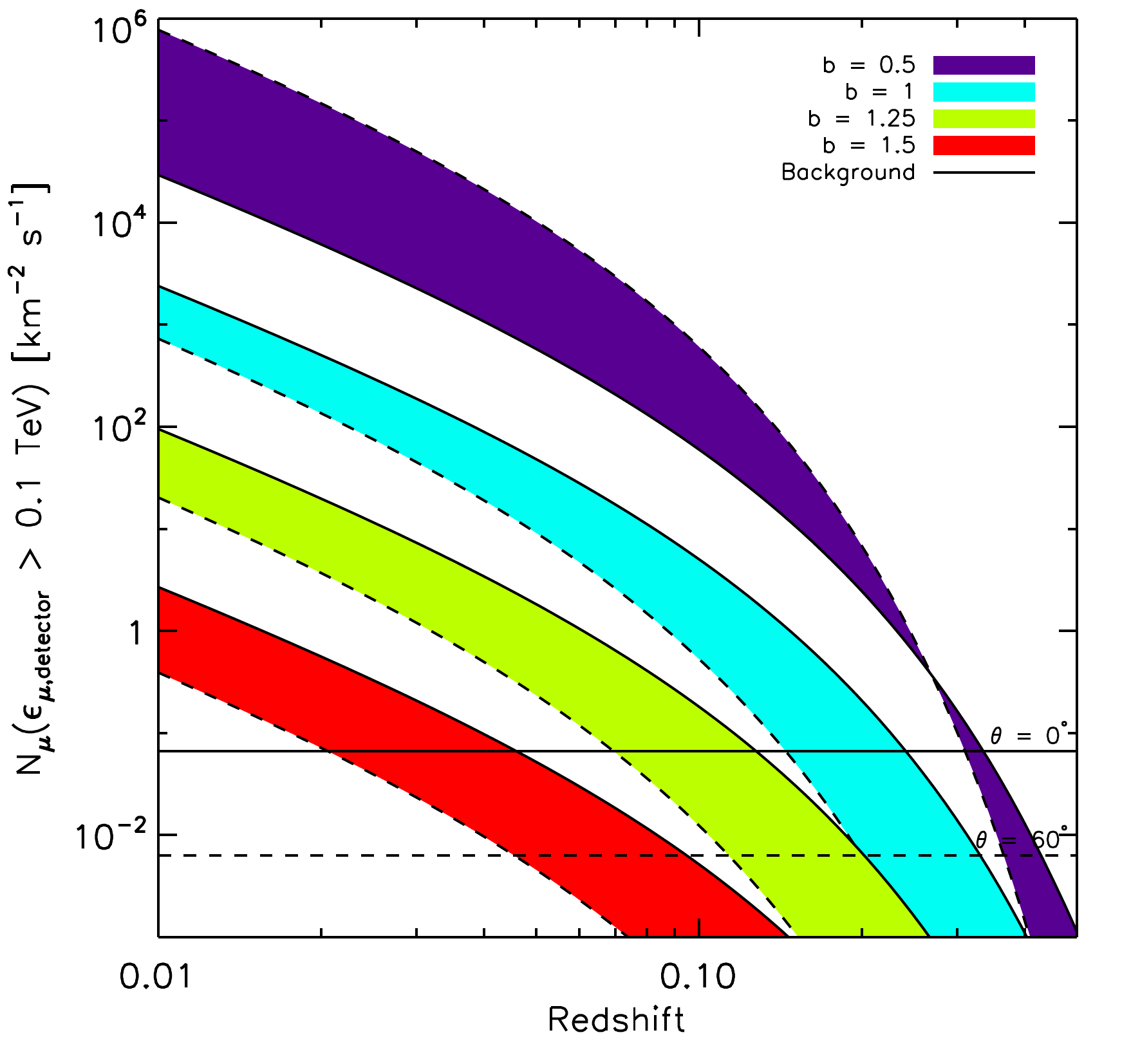}
   \caption{This figure plots the muon count with energies $\epsilon_{\mu,{\rm detector}} > 0.1 {\rm TeV}$ for GRB sources occuring at different redshifts. As in Figure \ref{fig:muonspec_detlevel_finke10}, dashed lines are for GRBs at zenith distance $\theta = 60^{\circ}$ while solid lines are for GRBs at $\theta = 0$. EBL attenuation is calculated using the model by \citet{fin10}.}
   \label{fig:muoncount_z_finke10}
\end{center}
\end{figure}
\begin{figure}
\begin{center}
   \includegraphics[width=84mm]{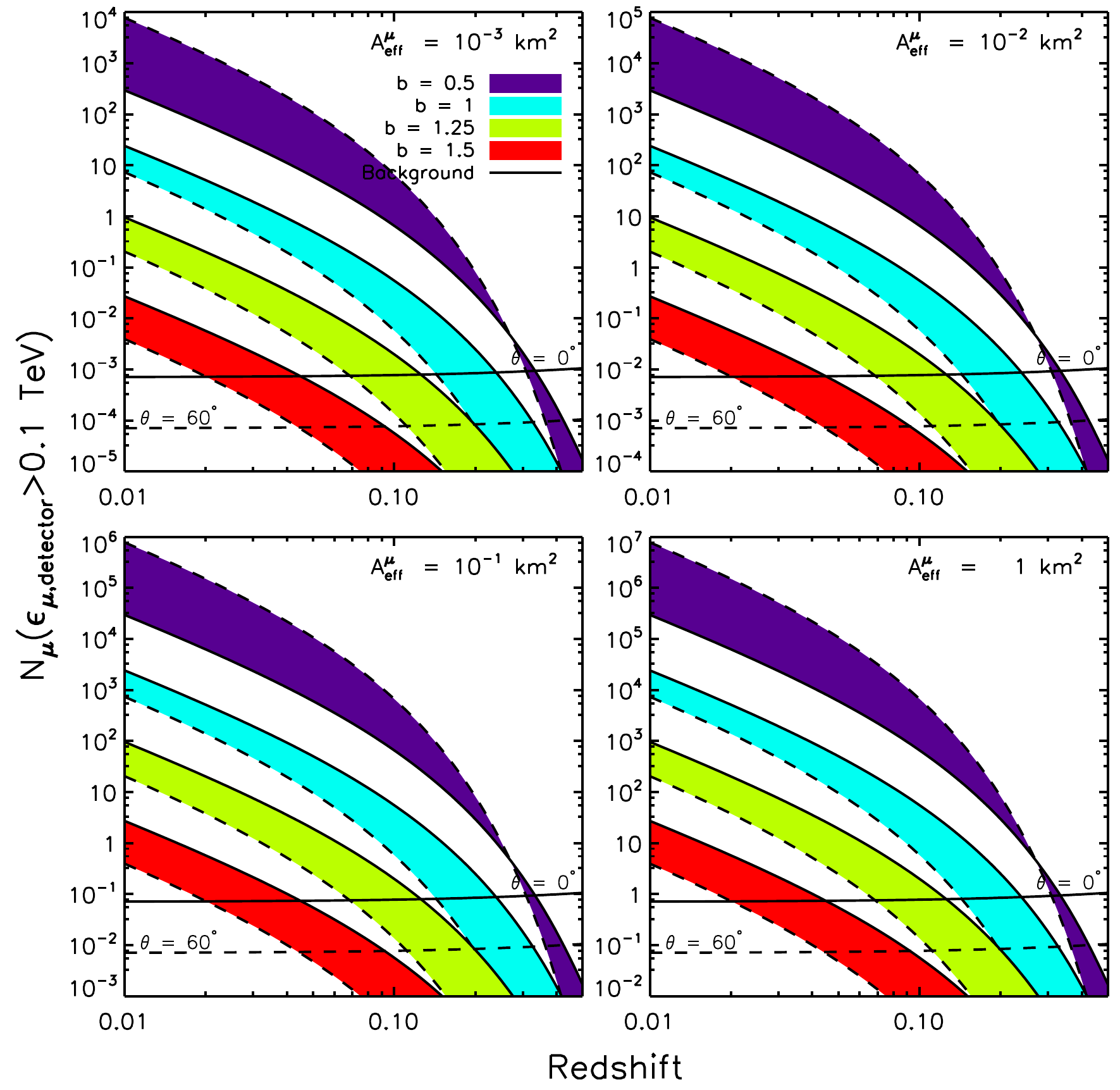}
   \caption{The total number of muons with energies $\epsilon_{\mu,{\rm detector}} > 0.1\; {\rm TeV}$ for GRBs from different redshifts and different spectral index as indicated by the color code in the legend. The intrinsic burst duration $\Delta t_*$ is assumed to be 10 sec, thus making $t_{90} = (1 + z)\Delta t_*$. The total muon count is calculated by assuming different detector size, which is also assumed to be independent of energy.}
   \label{fig:muoncount_aeff}
\end{center}
\end{figure}

The results shown in Figure \ref{fig:muonspec_detlevel_finke10} indicate that the number of muons reaching detector level depends heavily on the GRB's distance from us and its power spectrum. The redshift becomes the most important factor because it determines the number of photons that survives all the way from the GRB to the top of the atmosphere, the power spectrum comes second as it determines the number of photons originally produced in the GRB. 

The muon spectrum is then integrated to obtain a muon event rate with energies higher than $\epsilon_{\mu,{\rm detector}}$:
\begin{equation}
N_\mu(>\epsilon_{\mu,{\rm detector}}) = \int^{\infty}_{\epsilon_\mu,{\rm detector}} d\epsilon_{\mu} \frac{dN_{\mu}}{d\epsilon_\mu}
\end{equation}
The result of this integration is shown in Figure \ref{fig:muoncount_detlevel_finke10}, using the attenuation model by \citet{fin10}. This result can give us an idea of how many muon events per unit area per unit time can we expect from any GRB event with the given power spectrum, redshift, and zenith distance. 

To explore further the effect of distance to the muon event rate at detector level, in Figure \ref{fig:muoncount_z_finke10} the event rate of muons with energies higher than 0.1 TeV per unit area per unit time, $N_{\mu}(\epsilon_{\mu,{\rm detector}} > 0.1\;{\rm TeV})$, is plotted as a function of redshift. Here the black lines are the background rate from cosmic ray-induced muons at zenith distances $\theta = [0, 60^\circ]$ and it is independent of redshift.

Figure \ref{fig:muoncount_z_finke10} tells us the minimum redshift and maximum zenith distance to observe, for example, at least 1 muon event per kilometer square per second. For example, a GRB event with power spectrum $b=1.25$ that occurs at the zenith must have a redshift of $z\lesssim 0.07$ if we want to observe at least 1 muon event per kilometer square per seconds. The number of muons produced from a photon spectrum with $b=0.5$ and those from $b = 1.5$ exhibit a large discrepancies, ranging from $N_{\mu} \sim 1$ to $N_{\mu} \sim 10^{4}\;{\rm km}^{-2}\;{\rm s}^{-1}$. This is because a photon flux with a shallow spectrum can produce an electromagnetic shower that grows in the atmosphere, while flux with a steeper spectrum produce a shower that instead dissipate in the atmosphere.

The number of detectable muons depends also on the size of the Cherenkov detector. ANTARES is projected to have an effective muon area of $A^\mu_{\rm eff} \sim 10^{-2}\;{\rm km}^2$ while IceCube is expected to have an area the size of $A^\mu_{\rm eff} \sim 1\;{\rm km}^2$ \citep{hal09}. In Figure \ref{fig:muoncount_aeff}, I calculate the total number of detectable muons during the whole duration of the burst for four different detector size. The downgoing muon effective area is assumed to be $A^\mu_{\rm eff} = (10^{-3}, 10^{-2}, 0.1, 1)\; {\rm km}^2$, and is also assumed to be constant with respect to the muon energy. One quarter of Figure \ref{fig:muoncount_aeff} plots the total number of events for different detector sizes. 

We can see that although larger detector with $A^\mu_{\rm eff} = 1\;{\rm km}^2$ detect more unwanted background muons, they also see farther GRBs, up to $z\sim 0.3$ for $b = 0.5$. A detector the size of ANTARES, however, can only detect at least 1 muon event from a GRBs at redshift up to $z\sim 0.2$ for the same integral index.

\begin{figure}
\begin{center}
   \includegraphics[width=84mm]{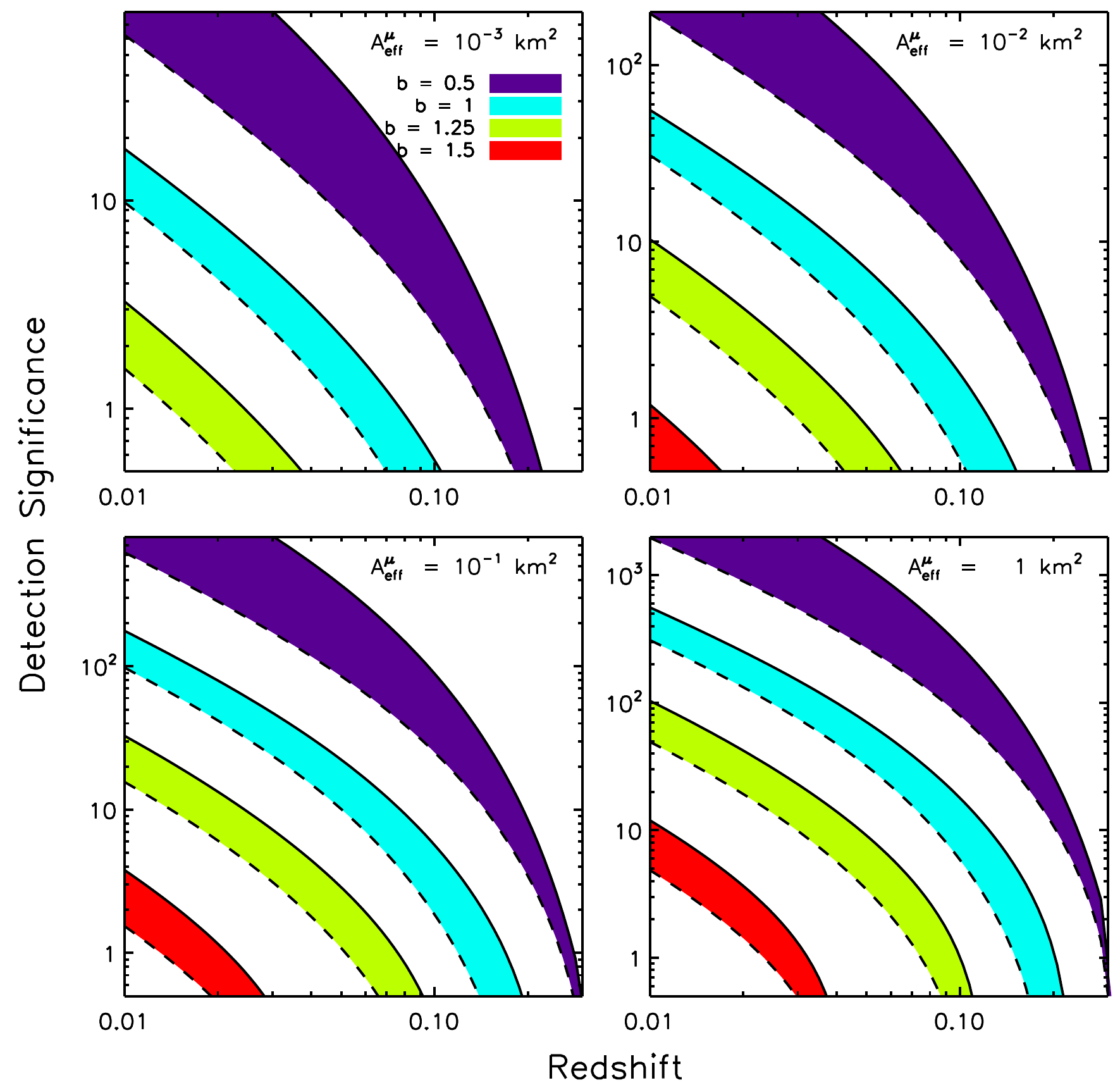}
   \caption{The muon signal detection significance for GRB sources located at different redshifts and different spectral index as indicated by the color code in the legend. The detection significance is calculated using the \citet{lima83} formula, and using the quantities shown in Figure \ref{fig:muoncount_aeff}.}
   \label{fig:signif_single}
\end{center}
\end{figure}

Since we know the number of signals and noises in our detector, we can now calculate the expected detection significance of each individual GRB as a function of redshift. The significance $S$ is calculated according to the procedure outlined by \citet{lima83}. The total signal $N_{\rm on}$ is the number of muon events within the $\theta_{\rm cone} = 1^\circ$ search cone and during the $t_{\rm on} = t_{90}$ time interval, while the total number of background $N_{\rm off}$ is the number of muon backgrounds within the same search area but some amount of time $t_{\rm off}$ before the GRB took place. The statistical significance $S$ (the number of standard deviation above background) is determined using the likelihood ratio method:

\begin{equation}
\begin{split}
S =&\; \sqrt{2}\left\{N_{\rm on}\ln\left[\frac{1 + \alpha}{\alpha}\left(\frac{N_{\rm on}}{N_{\rm on} + N_{\rm off}}\right)\right]\right. \\ &+ \left. N_{\rm off}\ln\left[(1 + \alpha)\left(\frac{N_{\rm off}}{N_{\rm on} + N_{\rm off}}\right)\right]\right\}^{1/2},
\end{split}
\end{equation}

where $\alpha$ is the ratio $\alpha\equiv t_{\rm on}/t_{\rm off}$. The time $t_{\rm off}$ to measure the background rate is taken to be 2 hours, i.e. $t_{\rm off} = 7200\;{\rm s}$, thus making $\alpha$ very low. The result of this calculation is shown in Figure \ref{fig:signif_single}, again with four different panels each assuming different detector sizes.

The Figure predicts the detection significance of observing GRBs with a certain power spectrum, zenith distance, and redshift. We can also use this result to determine the maximum redshift where a GRB has to occur if we want to have at least $3\sigma$ or $5\sigma$ detection significance. As an example, for an ANTARES-sized detector to detect a GRB signal with $5\sigma$ significance, a GRB event at zenith must be closer than $z\lesssim 0.05$ if its power spectrum is $b=1$.

\section{Conclusions}
\begin{figure}
\begin{center}
   \includegraphics[width=84mm]{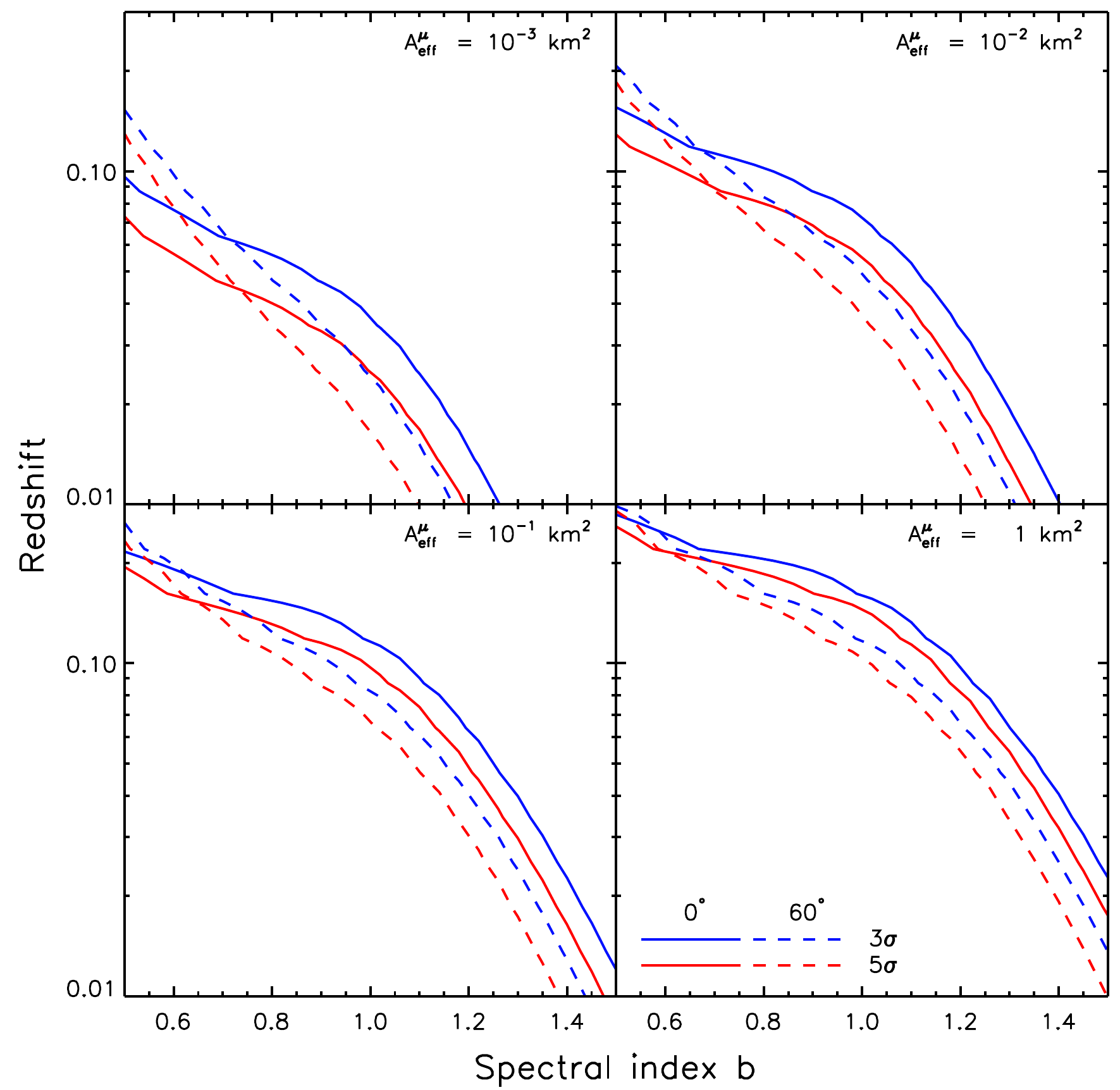}
   \caption{The combination of redshift $z$ and spectral intex $b$ that gives a detection significance of $3\sigma$ (blue lines) and $5\sigma$ (red lines), for GRB photons that came from zenith distances of $0^\circ$ (solid lines) or $60^\circ$ (dashed lines).}
   \label{fig:sig_bz}
\end{center}
\end{figure}
The two most important factors in detecting the TeV components of a GRB are its redshift and its spectral index. Its redshift determines the number of photons that survived to the top of the atmosphere, while the hardness of its spectrum determines whether the electromagnetic spectrum grows or dissipate in the atmosphere. The combination of these two values that give a detection significance of $3\sigma$ or $5\sigma$ over the background is presented in Figure \ref{fig:sig_bz}. A typical GRB has a spectral index $b = 1$--$1.25$ \citep{pre00, kan08}. For an ANTARES-type telescope, a typical GRB must then be located at least at redshift $z \lesssim 0.05$, while a larger telescope with a muon collecting area of $A^{\rm eff}_{\mu} = 1\;{\rm km}^2$ can see further up to $z\lesssim 0.1$ for a GRB with the same characteristic.

A recent analysis of {\em Fermi} GRB data by \citet{zha11} suggests that the peak of the distribution in $b$ has shifted instead to $b\sim 1.6$, a much steeper slope than what was suggested by previous observations. Consequently, the maximum redshift that permits a $3\sigma$ detection is lower: Redshift $z \lesssim 0.005$ for an ANTARES-type telescope and $z\lesssim 0.01$ for a ${\rm km}^3$ neutrino telescope. In the analysis of \citet{zha11}, the peak distribution of integral index $a$ is $a\sim -0.1$, which is not significantly different with previous results.

The limitation pertaining to distance proves to be a great hindrance to the detection of TeV $\gamma$-ray from GRBs, as there are not many GRBs with known redshift that took place at so close distance. Recent analysis of 425 {\it Swift} GRBs found out that the redshift distribution of GRBs is peaked at $z \sim 1$ \citep{but10}. Within this data set, there are 144 GRBs with known redshift and 3 of them have $z\leq 0.15$. This corresponds roughly to a fraction of $P(z\leq 0.15) \sim 7\times 10^{-3}$. 

From these results we can conclude that the secondary role of neutrino-telescopes as a gamma-ray telescope can only be played-out restrictively to the nearest GRB sources. As nearby GRBs tend to belong to a different population (i.e. short GRB) than the ones further away, further considerations must also be taken in view of the different luminosity and burst duration of this population.

The expected rate of muon signals calculated in this paper has not yet include the detection efficiency of the detector. To understand this detector effect a further study must be done, which require a Monte Carlo simulation of the interaction between the muon signals and the detector. This is a work reserved for future studies.

\section*{Acknowledgments}
I would like to thank Floyd Stecker and Rudy Gilmore for providing their optical depth tables that are essential for the calculations outlined in this paper. I also thank Ralph Wijers for carefully reading the manuscript and his numerous important suggestions, and also especially to Maarten de Jong from Nikhef, Amsterdam, for the impetus for this project, encouragement, and suggestions. I thank also the anonymous referee for the comments on the draft of this paper.

\bibliographystyle{mn2e_new}
\bibliography{thebibliography}
\label{lastpage}
\end{document}